\documentclass[letterpaper,12pt]{JHEP3}
\pdfoutput=1
\usepackage{amsmath}
\usepackage{amssymb}
\usepackage{bbm}
\usepackage{subfigure} 
\usepackage{epsfig}
\usepackage{yfonts}
\usepackage{graphicx}

\newcommand{\X}{\mathcal{X}}
\newcommand{\Z}{\mathcal{Z}}
\newcommand{\W}{\mathcal{W}}
\newcommand{\V}{\mathcal{V}}
\newcommand{\labell}[1]{\label{#1}}

\newcommand{\be}{\begin{equation}}
\newcommand{\ee}{\end{equation}}
\newcommand{\bea}{\begin{eqnarray}}
\newcommand{\eea}{\end{eqnarray}}
\newcommand{\ba}{\begin{eqnarray}}
\newcommand{\ea}{\end{eqnarray}}

\newcommand{\beq}{\begin{equation}}
\newcommand{\eeq}{\end{equation}}
\newcommand{\beqa}{\begin{eqnarray}}
\newcommand{\eeqa}{\end{eqnarray}}
\newcommand{\beqar}{\begin{eqnarray*}}
\newcommand{\eeqar}{\end{eqnarray*}}

\newcommand{\reef}[1]{(\ref{#1})}

\newcommand{\eg}{{\it e.g.,}\ }
\newcommand{\ie}{{\it i.e.,}\ }

\newcommand{\mt}[1]{\textrm{\tiny #1}}
\newcommand{\veps}{\varepsilon}

\newcommand{\la}{\lambda}
\newcommand{\lp}{\ell_{\mt P}}

\newcommand{\fin}{f_\infty}

\newcommand{\tL}{\tilde L}
\newcommand{\cL}{{\cal L}}

\def\L{{\mathcal L}}

\renewcommand{\a}[1]{a_\mt{#1}}
\renewcommand{\b}[1]{b_\mt{#1}}

\newcommand{\m}[1]{\mu_\mt{#1}}

\newcommand{\ads}{a_d^*}

\newcommand{\te}{t_\mt{E}}
\newcommand{\ct}{C_{T}} 

\preprint{arXiv:1011.5819 [hep-th]} 
\title{Holographic c-theorems in arbitrary dimensions}
\author{Robert C. Myers$^{1}$ and Aninda Sinha$^{1,2}$ \\
\it $^1$Perimeter Institute for Theoretical Physics\\
\ Waterloo, Ontario N2L 2Y5, Canada\\
$^2$Centre for High Energy Physics, Indian Institute of Science,\\
\ C.V.Raman Avenue, Bangalore 560012, India}

\abstract{We re-examine holographic versions of the c-theorem and
entanglement entropy in the context of higher curvature gravity and the
AdS/CFT correspondence. We select the gravity theories by tuning the
gravitational couplings to eliminate non-unitary operators in the boundary
theory and demonstrate that all of these theories obey a holographic c-theorem.
In cases where the dual CFT is even-dimensional, we show that the
quantity that flows is the central charge associated with the A-type
trace anomaly. Here, unlike in conventional holographic constructions
with Einstein gravity, we are able to distinguish this quantity from other
central charges or the leading coefficient in the entropy density of
a thermal bath. In general,
we are also able to identify this quantity with the coefficient of a
universal contribution to the entanglement entropy in a particular
construction. Our results suggest that these coefficients appearing in
entanglement entropy play the role of central charges in odd-dimensional
CFT's. We conjecture a new c-theorem on the space of odd-dimensional
field theories, which extends Cardy's proposal for even dimensions.
Beyond holography, we were able
to show that for any even-dimensional CFT, the universal coefficient
appearing the entanglement entropy which we calculate is precisely the
A-type central charge.}

\begin{document}

\section{Introduction}

Zamolodchikov's c-theorem \cite{zam} states that for quantum field
theories (QFT's) in two dimensions, there exists a positive definite
real function $c(g)$ on the space of couplings $g^i$, which satisfies
the following three properties: {\it i)} $c(g)$ is monotonically
decreasing along renormalization group (RG) flows. {\it ii)} $c(g)$ is
stationary at RG fixed points $g^i=(g^*)^i$, \ie $\partial
c(g)/\partial g^i|_{g^*}=0$. {\it iii)} At RG fixed points, $c(g)$
equals the central charge of the corresponding conformal field theory
(CFT). This remarkable result requires only very simple conditions of
the QFT's with the proof relying only on the Euclidean group of
symmetries, the existence of a conserved stress-energy tensor and
unitarity in the field theory. A direct consequence of the c-theorem is
that in any renormalization group (RG) flow connecting two fixed
points,
\be (c)_{\rm UV}\ge (c)_{\rm IR}\,. \labell{beta} \ee
That is, the central charge of  the CFT describing the ultraviolet
fixed point is larger than (or equal to) that at the infrared fixed
point. An intuitive understanding of this result comes from the
interpretation that the central charge provides a measure of the number
of degrees of freedom of the underlying CFT. Its decrease along the RG
flow can then be seen as a consequence of integrating out high-energy
degrees of freedom in the Wilsonian approach to the renormalization
group.

There have been various suggestions on how such a result might extend
to quantum field theories in higher $d$. One approach refers to the
trace anomaly which is fixed by the central charge $c$ for
two-dimensional CFT's \cite{traca}, \ie
 \be
\langle\, T^a{}_a\, \rangle =-\frac{c}{12} R\,. \labell{trace2}
 \ee
Turning to $d=4$, this expression generalizes to \cite{traca}
 \be
\langle\, T^a{}_a \rangle=\frac{c}{16\pi^2}
W_{abcd}W^{abcd}-\frac{a}{16\pi^2}\left( R_{abcd}R^{abcd}-4R_{ab}R^{ab}
+ R^2\right) -\frac{a'}{16\pi^2} \nabla^2 R\,, \labell{trace4}
 \ee
where $W_{abcd}W^{abcd}= R_{abcd}R^{abcd}-2R_{ab}R^{ab}+R^2/3$ is the 
square of the four-dimensional Weyl tensor and the expression in the
second term is proportional to the four-dimensional Euler density.
Hence, the question naturally arises: do any of $c$, $a$ or $a'$
satisfy a c-theorem under RG flows? Since $a'$ is scheme dependent
\cite{schema} and cannot be defined globally \cite{global}, it is not a
useful charge to consider. With regards to the four-dimensional central
charge $c$, many counter-examples are now known where eq.~\reef{beta}
is not satisfied \cite{count1,anselmi}. Further, the latter
investigations \cite{anselmi} also demonstrated that any linear
combination of $c$ and $a$ will not generically satisfy
eq.~\reef{beta}. This leaves us only to consider the central charge
$a$.

It was Cardy \cite{cardy} who originally conjectured that $a$ should
decrease monotonically along RG flows of four-dimensional QFT's.
Numerous nontrivial examples have been found supporting this
conjecture, including perturbative fixed points \cite{jack} and
supersymmetric gauge theories \cite{anselmi,ken}. While a
counter-example to Cardy's conjecture in $d=4$ was proposed in
\cite{yuji}, recently, a certain flaw in this analysis was identified
and so this possible counter-example is removed \cite{yuji2}. Further,
as we review below, support for such a c-theorem in $d=4$ was found
with the AdS/CFT correspondence \cite{gubser,friends} -- however, in
the class of holographic theories studied there, $c=a$ and so both $a$
and $c$ satisfy eq.~\reef{beta}. While Cardy's conjecture is supported
by numerous nontrivial examples, a general proof is still lacking. A
primary purpose of this paper is to report further evidence for the
conjecture coming from a broader class of holographic models, where in
particular the central charges $a$ and $c$ are distinct. An preliminary
report of these results was given in \cite{new}.

Cardy's conjecture \cite{cardy} actually referred to any even number of
spacetime dimensions. For even $d$, the trace anomaly for CFT's in a
curved background can be written as \cite{traca}
\be \langle\,T^a{}_a\,\rangle = \sum B_i\,I_i -2\,(-)^{d/2}A\, E_d +
B'\,\nabla_a J^a \labell{trace} \ee
where $E_d$ is the Euler density in $d$ dimensions and $I_i$ are the
independent Weyl invariants of weight $-d$.\footnote{A note on our
conventions: The stress tensor is defined by
$T_{ab}\equiv-2/\sqrt{-g}\,\delta I/\delta g^{ab}$. The Euler density
$E_d$ is normalized so that on a $d$-dimensional sphere:
$\oint_{S^d}d^d\!x\sqrt{g}\, E_d =2$. The Weyl invariants $I_i$ are
constructed from contractions of $d/2$ curvatures or $d/2-k$ curvatures
and $2k$ covariant derivatives. There is some ambiguity in the
construction of the $I_i$ which we (partially) fix by demanding that
these invariants vanish when evaluated on a round $d$-dimensional
sphere, \ie $I_i|_{S^d}=0$.} Finally the last term is a conformally
invariant but also scheme-dependent total derivative. That is, this
last contribution can be changed or even eliminated by adding a
(finite) covariant counter-term to the action. Cardy's proposal is then
that, in any even $d$, for RG flows connecting two fixed points
\be (A)_{\rm UV}\ge (A)_{\rm IR}\,. \labell{beta2} \ee
Of course, this coincides with Zamolodchikov's result in $d=2$ where
$A= c/12$ and it matches the above discussion for $d=4$ where $A=a$
with our present choice of normalization.$^{1,}$\footnote{Further, in
comparing eqs.~\reef{trace4} and \reef{trace}, we see for $d=4$ there
is a single invariant $I_1$ corresponding to the Weyl tensor squared
and for which $c=16\pi^2B_1$.}

One of the advantages of the investigating RG flows in a holographic
framework is that the results are readily extended to arbitrary
dimensions \cite{gubser,new}. In \cite{new} and in the following, we
examine holographic models with higher curvature gravity in the
($d$+1)-dimensional bulk, which allows us to distinguish the central
charges appearing in the trace anomaly \reef{trace} of the
$d$-dimensional boundary CFT's for even $d$. Hence we are able to
discriminate between the behaviour of the various central charges in RG
flows and we find that only $A$ has a natural monotonic flow, giving
further support for Cardy's conjecture \reef{beta2}. Our analysis of
holographic RG flows applies in arbitrary higher $d$ and in fact, we
find a certain quantity, denoted $\ads$, satisfying an inequality such
as that given in eq.~\reef{beta} or \reef{beta2} for any $d$, that is,
for both even and odd $d$. There is no trace anomaly for odd $d$ and so
some new interpretation for $\ads$ must be found in this case.
Following \cite{new}, we identify this quantity with the coefficient of
a universal contribution to the entanglement entropy for a particular
construction for both odd and even $d$. Our results suggest that these
coefficients appearing in entanglement entropy play the role of central
charges for odd-dimensional CFT's and allow us to conjecture a
c-theorem on the space of odd-dimensional field theories. However, we
must emphasize that our higher curvature gravity actions are not
derived from string theory. These theories should be regarded as toy
models which allow us to explore of the role of higher curvature terms
in holography. Ultimately, one would like to develop a better
understanding of string theory in order study interesting holographic
backgrounds with high curvatures and study the possibility of a
holographic c-theorem in this framework.

An overview of the paper is as follows: We begin with a review of
holographic c-theorem for Einstein gravity \cite{gubser,friends} in
section \ref{one}. Further, we describe a similar c-theorem for a
particular higher curvature theory, known as quasi-topological gravity
\cite{old1,old2}. In section \ref{two}, we extend our discussion to
gravity theories with more general higher curvature interactions. We
propose that the couplings of these new interactions should be tuned to
remove any non-unitary operators from the dual boundary theory. With
this constraint, we find that the resulting theories automatically
satisfy a holographic c-theorem. In both of these sections, we are able
to show that for even $d$, the quantity $\ads$ which obeys the
holographic c-theorem is precisely the central charge $A$ in
eq.~\reef{trace}. In section \ref{EE1}, we demonstrate that $\ads$
appears in a  certain calculation of entanglement entropy for odd or
even $d$. In particular, we place the $d$-dimensional CFT on
$S^{d-1}\times R$ and calculating the entanglement entropy of the
ground state between two halves of the sphere. We then find a universal
contribution: $S_{univ}\propto \ads$. In this section, the entanglement
entropy is calculated by relating it to the thermal entropy of the CFT
on the hyperbolic plane, \ie $H^{d-1}\times R$, with a particular
temperature. In section \ref{EE2}, we describe a more conventional
calculation of the entanglement entropy in the boundary CFT using the
replica trick. The latter is translated to a holographic calculation in
the bulk gravity theory and we reproduce the same results as in the
previous section. Here we note that these calculations are distinct
from the standard holographic calculations of entanglement entropy
\cite{taka1,taka2}, which are only applicable with the bulk theory is
Einstein gravity. In section \ref{EEcft}, we also extend our
calculation using the replica trick to any CFT in even dimensions and
show (without holography) that in general the universal coefficient is
precisely the central charge $A$. Returning to the holographic
framework in section \ref{density}, we show that $\ads$ can be thought
of as counting the degrees of freedom in the boundary CFT. In section
\ref{other}, we compare our results with two other proposals for
charges which may satisfy a c-theorem in higher dimensions. We
explicitly show that $\ads$ does not correspond to the coefficient
governing the leading singularity of the two-point function of the
stress tensor or the leading coefficient in the entropy density of a
thermal bath. Further, we show that the latter coefficient need not
satisfy a holographic c-theorem within the class of boundary theories
described by quasi-topological gravity. We conclude with a discussion
of our results and possible future directions in section \ref{discuss}.
We also have some appendices containing related calculations. In
appendix \ref{aflow}, we examine the holographic RG flows more
generally for the theories considered in section \ref{one}. Appendix
\ref{ageneral} makes some preliminary comments on establishing a
holographic c-theorem when the matter fields couple to the higher
curvature interactions. Finally in appendix \ref{anonrel}, we make some
brief comments about possible c-theorems for holographic models with
the non-relativistic Schr\"odinger symmetry.

\section{Holographic c-theorems -- Take One} \label{one}

The c-theorem was first considered in the context of the AdS/CFT
correspondence by \cite{gubser,friends}. There one begins with
($d$+1)-dimensional Einstein gravity coupled to various matter fields:
 \be I=\frac{1}{2\lp^{d-1}}\int d^{d+1} x \, \sqrt{-g} \left(R+
{\mathcal L}_{\rm matter}\right)
 \labell{act0}
 \ee
The matter theory is assumed to have various stationary points where
${\mathcal L}_{\rm matter}={d(d-1)\alpha_i/L^2}$ with some canonical
scale $L$. While the latter is phrased in a general way, it is useful
to keep in mind a simple example in the following. Namely, the matter
sector here could naturally be a scalar field theory where the
potential has a number of extrema with
$V_{i,crit}=-{d(d-1)\alpha_i/L^2}$. The vacuum energy or cosmological
constant is negative at all of the relevant stationary points and it is
a convenient notation to introduce $\alpha_i$ to distinguish the
different values. At these points, the vacuum solution for the Einstein
theory is simply AdS$_{d+1}$ with the curvature scale given by $\tilde
L^2=L^2/\alpha_i$.

Next one considers solutions of the above theory \reef{act0} where the
scalar sits at one fixed point in the asymptotic (UV) region and makes
a smooth transition to another fixed point in the interior (IR) region.
Such a solution can be interpreted as a holographic representation of a
renormalization group flow, in which the boundary CFT flows from one
fixed point in the UV to another in the IR. The spacetime geometry for
these holographic RG flows is conveniently described with a metric of
the form
\be ds^2=e^{2 A(r)}\left( -dt^2+ d\vec{x}_{d-1}^2 \right)+dr^2\,.
\labell{metric} \ee
This metric becomes that for AdS$_{d+1}$ with $A(r)=r/\tilde L$ at the
stationary points. Now one defines \cite{gubser}:
\be a(r)\equiv\frac{\pi^{d/2}}{\Gamma\left(d/2\right)\left(\lp
A'(r)\right)^{d-1}}\,, \labell{def0} \ee
where `prime' denotes a derivative with respect to $r$. Then for
general solutions with the above metric \reef{metric}, one finds
\bea a'(r)&=&-\frac{(d-1)\pi^{d/2}}{\Gamma\left(d/2\right)\lp^{d-1}
A'(r)^d} A''(r)
\labell{magic0}\\
&=& -\frac{\pi^{d/2}}{\Gamma\left(d/2\right)\lp^{d-1}
A'(r)^d}\left(T^t{}_t-T^r{}_r\right) \ge0\,.\nonumber
 \eea
In the second equality above, the Einstein equations are used to
eliminate $A''(r)$ in favour of components of the stress tensor. The
final inequality assumes that the matter fields obey the null energy
condition \cite{HE}. Combining this monotonic evolution of $a(r)$ with
$r$ with the standard connection between $r$ and energy scale in the
CFT, $a(r)$ always decreases in flowing from the UV (large $r$) to the
IR (small $r$).

To make a more precise interpretation of the bulk solutions in terms of
the boundary CFT, it is simplest to focus the discussion on $d=4$ at
this point. In this case, the holographic trace anomaly \cite{sken}
allows one to calculate the two central charges, $a$ and $c$, of the
four-dimensional CFT -- see eq.~\reef{trace4}. For any of the AdS$_5$
vacua with a curvature scale $\tilde L$, one finds
\be c=a \qquad {\rm and}\qquad a=\pi^2\,\frac{\tilde
L^3}{\lp^3}=a(r)\big|_{AdS}\,. \labell{acharge} \ee
As we emphasize below, the equality of the two central charges results
because the bulk theory is Einstein gravity \cite{sken}. However, the
important observation is that the value of the flow function
\reef{def0} will precisely match that of the central charges in the
dual CFT at each of the fixed points. Hence with the assumption of the
null energy condition, the holographic CFT's dual to Einstein gravity
\reef{action} satisfy Cardy's proposed c-theorem. That is, for these
holographic RG flows, $a$ is always larger at the UV fixed point than
at the IR fixed point. Of course, these holographic models do not
distinguish between the flow of $a$ and $c$, and so the central charge
$c$ obeys the same inequality as well.

It has long been known that to construct a holographic model where
$a\ne c$, the gravity action must include higher curvature interactions
\cite{highc}. In part, this motivated the recent construction of the
higher curvature theory, known as quasi-topological gravity
\cite{old1}. This gravitational theory should be regarded as a toy
model which allows us to explore the behaviour of a broader class of
holographic CFT's. It was demonstrated in \cite{new} that this bulk
theory also naturally exhibits a holographic c-theorem, as follows: The
action for quasi-topological gravity can be written as \cite{old1}
\beqa
 I &=& \frac{1}{2\lp^{d-1}} \int \mathrm{d}^{d+1}x \, \sqrt{-g}\, \left[
\frac{d(d-1)}{L^2}\alpha + R + \frac{\lambda L^2}{(d-2)(d-3)}\X_4
\right.
 \nonumber\\
&&\qquad\qquad\quad\left.-\frac{8(2d-1)\,\mu L^4
}{(d-5)(d-2)(3d^2-21d+4)}\,\Z_{d+1}\right] \labell{ActD}
 \eeqa
where $\X_4$ is the four-dimensional Euler density, as used in
Gauss-Bonnet gravity \cite{lovel},
 \beq
\X_4=R_{abcd}R^{abcd}-4\,R_{ab}R^{ab}+R^2 \labell{GBterm}
 \eeq
and $\Z_{d+1}$ is the new curvature-cubed interaction \cite{old1,chile}
 \beqa \Z_{d+1}&=& R_a{}^c{}_b{}^{d} R_c{}^e{}_d{}^{f}R_e{}^a{}_f{}^{b}
 +
\frac{1}{(2d-1)(d-3)}\left(\frac{3(3d-5)}{8}R_{a b c d}R^{a b c d}
R \right.\nonumber\\
&&\quad-\,3(d-1) R_{a b c d}R^{a b c}{}_{e}R^{d e}+ 3(d-1)R_{a b c d}
R^{a c}R^{b d}\labell{result}\\
&&\quad\left.+\,6(d-1)R_a{}^{b}R_b{}^{c}R_c{}^{a}-\frac{3(3d-1)}{2}
R_a{}^{b}R_b{}^{a}R +\frac{3(d-1)}{8}R^3\right)\,.\nonumber
 \eeqa
This action is written for any (boundary) dimension $d\ge4$, although
we set $\mu=0$ for $d=5$ to avoid the singular behaviour of the
pre-factor of $\Z_{d+1}$ in eq.~\reef{ActD}. By introducing
interactions quadratic and cubic in the curvature, this holographic
model allows one to explore the full three-parameter space of
coefficients controlling the two- and three-point functions of the
stress tensor in a general $d$-dimensional CFT \cite{osborn}. The
reader should keep in mind that this action \reef{ActD} was not derived
from string theory. Rather, as noted above, it was constructed as a toy
model to allow us to explore a broader class of holographic CFT's while
maintaining control within the gravity calculations. However, we should
also note that the gravitational couplings, $\lambda$ and $\mu$, in
eq.~\reef{ActD} are constrained to be not very large, otherwise one
finds that the dual CFT is inconsistent -- for a more precise
discussion, see \cite{old2,consist,holoGB,fate}. We emphasize that the
discussions here and in \cite{old1,old2} should only be regarded as an
initial exploration of the role of higher curvature terms in
holography. Ultimately, one would like to develop our understanding of
string theory to the point where we can study interesting holographic
backgrounds with high curvatures.

We have also introduced a factor of $\alpha(>0)$ in the cosmological
constant term above in anticipation of our consideration of holographic
RG flows below. The idea is that as in eq.~\reef{ActD}, the gravity
theory is coupled to a standard matter theory, \eg a scalar field, with
various stationary points which yield different values for the
parameter $\alpha$. At any of these stationary points, there is an
AdS$_{d+1}$ solution with a curvature scale $\tilde L^2=L^2/\fin$ where
\be \alpha=\fin-\lambda\,\fin^2-\mu\,\fin^2\,. \labell{cubic} \ee
In general, this equation yields three roots for $\fin$. However, for
any choice of the couplings $\la$ and $\mu$, at most one of these roots
corresponds to a ghost-free AdS vacuum which supports nonsingular black
hole solutions, as described in detail in \cite{old1}. Further, in the
case that the couplings, $\lambda$ and $\mu$, are not large, this will
be the root that is continuously connected to the single root (\ie
$\fin=\alpha$) that remains in the limit $\lambda,\,\mu\rightarrow 0$.
Implicitly, we will be working in this regime of the coupling space and
with this particular root in the following.

Originally, this action \reef{ActD} was constructed to give a theory
for which AdS black hole solutions could be easily found analytically
\cite{old1}. However, another remarkable property of quasi-topological
gravity is that the linearized graviton equations in the AdS$_{d+1}$
vacuum are only second order in derivatives \cite{old1}. In fact, up to
an overall numerical factor, the linearized equations are precisely the
same as those for Einstein gravity in the AdS$_{d+1}$ background. If we
focus on $d=4$ for a moment, the techniques of \cite{sken} can be
applied to calculate the central charges \cite{old2}
 \beqa
c&=& \pi^2\frac{\tL^3}{\lp^3}\left(1-2\la\fin -3\mu\fin^2 \right)\ ,
 \labell{cc}\\
a&=&\pi^2\frac{\tL^3}{\lp^3}\left(1-6\la\fin +9\mu\fin^2 \right)\ .
 \labell{aa}
 \eeqa
These expressions make clear that $a\ne c$ as long as the higher
curvature couplings are nonvanishing.

Now to examine holographic RG flows in quasi-topological gravity (for
arbitrary $d$), we adopt the same metric ansatz \reef{metric}. As
above, at a stationary point of the matter sector with a fixed
$\alpha$, the $AdS_{d+1}$ vacua again correspond to $A(r)=r/\tilde L$.
In this case, we construct a new flow function as \cite{new}
 \bea
a_d(r)&\equiv&{\pi^{d/2}\over\Gamma\left(d/2\right)
\left(\lp A'(r)\right)^{d-1}}\, \labell{adfun}\\
&&\times \
 \left(1-{2(d-1)\over
d-3}\lambda L^2A'(r)^2
- {3(d-1)\over d-5}\mu L^4A'(r)^4 \right)\,. \nonumber \eea
Now examining the radial evolution of $a_d(r)$, we find
 \bea
a'(r)&=&-\frac{(d-1)\pi^{d/2}}{\Gamma\left(d/2\right)\lp^{d-1} A'(r)^d}
A''(r) \left(1-{2}\lambda L^2A'(r)^2 - {3}\mu L^4A'(r)^4 \right)
\labell{magic2}\\
&=& -\frac{\pi^{d/2}}{\Gamma\left(d/2\right)\lp^{d-1}
A'(r)^d}\left(T^t{}_t-T^r{}_r\right) \ge0\,.\nonumber
 \eea
Again, the gravitational equations of motion allow us to introduce the
components of the stress tensor in going from the first to second line.
Further, as above, we also assume the null energy condition for the
final inequality to hold. We apply this constraint here in the spirit
of constructing a toy model with reasonable physical properties. One
might note that violations of the null energy condition have been
argued to lead to instabilities quite generally \cite{null}. In the
context of a full string theory or theory of quantum gravity, we expect
that this condition will be relaxed but to make progress here, we
assume that the matter sector continues to obey the null energy
condition as a pragmatic choice -- see section \ref{discuss} for
further discussion. With the latter assumption then, $a(r)$ evolves
monotonically along the holographic RG flows and we can conclude that
the corresponding `central charge' is always larger in the UV than at
the IR fixed point.

Note that there is a technical point which we must address for odd $d$.
In this case, it could be that the expression in the second line of
eq.~\reef{magic2} is negative if $A'(r)<0$. However, we can rule out
this possibility as follows: By construction, our flow geometry will
have an asymptotically AdS$_{d+1}$ region at large $r$ where $A'(r)>0$.
Now imagine that in the interior, $A'(r)$ is negative over some region
$r_0<r<r_1$ and positive from $r_1$ out to the asymptotic boundary.
Hence at the radius $r_1$, we must have had $A'(r_1)=0$ and
$A''(r_1)>0$. However, this leads to a contradiction. If we evaluate
the equation of motion $(d-1)A''(r)\,(1-2\lambda L^2 A'(r)^2-3\mu L^4
A'(r)^4)=T_t^t-T_r^r$ at $r=r_1$ and combine this result with the null
energy condition, we find $A''(r_1)\le0$. Hence our assumption that
there is some region where $A'(r)<0$ must be incorrect.\footnote{In the
special case that $A''(r_1)=0$, we can assume that it can be expressed
in terms of a Taylor expansion around $r_1$. To leading order, we would
have $A''(r)\simeq k\,(r-r_1)^n$ with $k>0$ and $n$ is some positive
and even integer. Then, as above, we again find a contradiction in the
vicinity around $r=r_1$ and reach the same conclusion.}

Let us denote the fixed point value of the flow function \reef{adfun}
as
 \be
\ads\equiv a_d(r)|_{AdS}={\pi^{d/2}\tilde
L^{d-1}\over\Gamma\left(d/2\right)\lp^{d-1}} \left(1-{2(d-1)\over
d-3}\lambda \fin - {3(d-1)\over d-5}\mu \fin^2 \right)\,.
 \labell{astar}
 \ee
Then with eq.~\reef{magic2}, our holographic model satisfies a
holographic c-theorem which specifies that
\be \left(\ads\right)_{UV}\ge \left(\ads\right)_{IR}\,. \labell{beta3}
\ee
Having found that $\ads$ satisfies a c-theorem, one is left to
determine what this quantity corresponds to in the dual CFT. Inserting
$d=4$ into eq.~\reef{astar} and comparing with eq.~\reef{aa}, we see
that $a_4^*$ is precisely the central charge $a$. Motivated by Cardy's
conjecture \reef{beta2} for a c-theorem in QFT's in even dimensional
spacetimes, it is natural to compare $\ads$ to the coefficient $A$ in
eq.~\reef{trace}. In fact, using the approach of
\cite{adam},\footnote{See discussion around eq.~\reef{adamA} for more
details of this calculation.} one readily confirms that there is again
a precise match
\be \ads =  A \quad{\rm for\ even\ }d\,. \labell{evend} \ee
Hence again, we find support for Cardy's conjecture with this broad
class of holographic CFT's. However, we must seek a broader definition
of $\ads$ in order to understand our results for odd $d$. We address
this question in sections \ref{EE1} and \ref{EE2}, where we show that
$\ads$ emerges in a certain calculation of entanglement entropy.

\section{Holographic c-theorems -- Take Two} \label{two}

In section \ref{one}, we have shown that quasi-topological gravity
naturally gives rise to a holographic c-theorem. Further at the fixed
points, we identified the quantity that decreases along the RG flows as
the coefficient of the A-type trace anomaly of the dual conformal field
theory, for even $d$. Now we would like to test how robust this result
is by expanding our considerations of holographic RG flows to a broader
class of gravitational theories. In the following, we begin with a
completely general curvature-cubed action and develop a series of
constraints so that the resulting holographic model is physically
reasonable. Again, our analysis here should be considered as an
exploration of holography with certain toy models which display
credible physical properties.

To begin, we will consider an action of the form
 \be
I=\frac{1}{2\lp^{d-1}}\int\mathrm{d}^{d+1}x \, \sqrt{-g}\, \left[
\frac{d(d-1)}{L^2}\alpha+R+ L^2 \widetilde\X+ L^4 \widetilde\Z\right]
\labell{action}
 \ee
where $\X$ and $\Z$ contain general interactions quadratic and cubic in
the curvature
 \beqa
 \widetilde\X&=&\lambda_1 R_{abcd}R^{abcd}+\la_2 R_{ab}R^{ab}+\la_3 R^2\ ,
 \label{euler4}\\
 \widetilde\Z&=&  \mu_1\, R_{a\,\,b}^{\,\,c\,\,\,d}
R_{c\,\,d}^{\,\,e\,\,\,f} R_{e\,\,f}^{\,\,a\,\,\,b}+ \mu_2\,
R_{ab}^{\,\,\,\,\,\,cd} R_{cd}^{\,\,\,\,\,\,ef} R_{ef}^{\,\,\,\,\,\,
ab} +\mu_3\, R_{a b c d} R^{a b c}_{\,\,\,\,\,\,\,e} R^{d e}
 \labell{ZZ}\\
&&\quad
  +\,
\mu_4\, R_{a b c d} R^{a b c d} R+ \mu_5\,R_{a b c d} R^{a c}R^{b d}
 +\mu_6\, R_a^{\,\,b}
R_b^{\,\,c} R_c^{\,\,a}+ \mu_7\, R_a^{\,\,b} R_b^{\,\,a} R +\, \mu_8\,
R^3\,.\nonumber
 \eeqa
In constructing $\widetilde\Z$, we began with all possible
six-derivative interactions and eliminated terms which are redundant
because of index symmetries or the Bianchi identities or which are
total derivatives \cite{old1}. In fact, this construction yields two
additional independent terms, $\nabla_a R_{bc} \nabla^aR^{bc}$ and
$\nabla_a R\, \nabla^a R$. However, we have discarded these terms in
eq.~\reef{ZZ} for simplicity as we would find that the corresponding
coefficients are always set to zero with the constraints introduced in
the following discussion.

We again assume that there is a matter sector, which exhibits
stationary points with different values of $\alpha$ in the cosmological
constant term in eq.~\reef{action}. At any of these critical points,
there is an AdS$_{d+1}$ with a curvature scale $\tilde L^2=L^2/\fin$
where
 \be
\alpha=\fin-\hat\lambda\,\fin^2-\hat\mu\,\fin^2\,,
 \labell{cubic9}
 \ee
with
 \bea
\hat\lambda&=&\frac{d-3}{d-1}\Big(2\,\lambda_1+d\,\lambda_2+
d(d+1)\,\lambda_3\Big)\,,
 \nonumber\\
\hat\mu&=&-\frac{d-5}{d-1}\Big((d-1)\,\mu_1+4\,\mu_2
+2d\,\mu_3+2d(d+1)\,\mu_4
 \labell{cubic8}\\
 &&\qquad\qquad +d^2\,\mu_5+d^2\,\mu_6+d^2(d+1)\,
\mu_7+d^2(d+1)^2\mu_8\Big)\,.
 \nonumber
 \eea
Of course, we have arranged eq.~\reef{cubic9} to take the same form as
eq.~\reef{cubic} in the previous section. In general, this cubic
equation again yields three roots for $\fin$. However, as in the
previous section, when the couplings, $\lambda_i$ and $\mu_i$, are not
large, there will be one root that is continuously connected to the
single root (\ie $\fin=\alpha$) that remains in the limit of Einstein
gravity, \ie $\lambda_i,\,\mu_i\rightarrow 0$. Implicitly, we will be
working in this regime in the following and $\fin$ will refer to this
particular root. We have not analyzed the general theory in great
detail but we expect that, as for quasi-topological gravity, the vacua
corresponding to any other (real) roots will be problematic
\cite{old1}.

While we could examine holographic RG flows with this action with
general curvature-squared and -cubed interactions, it seems
unreasonable to expect that any such arbitrary gravity theory should
yield a holographic c-theorem, just as it is unreasonable to expect
that any arbitrary quantum field theories should satisfy a c-theorem.
In particular, we do not expect that non-unitary QFT's will satisfy a
c-theorem. Hence we must ask how should we constrain the new couplings
in this gravitational action \reef{action} in order to produce a
physically credible model. Quasi-topological gravity has a number of
interesting properties which make it a reasonable toy model for
holographic studies. One striking feature of the theory was that
although the general equations of motion are fourth order in
derivatives, if the equations of motion for gravitons propagating in
the AdS vacuum are only second order \cite{old1}. While this feature
greatly facilitates holographic investigations of this theory, as
explicitly seen in \cite{old2}, there is a deeper significance to this
property, as we now discuss.

Given the general gravitational action \reef{action}, the full
equations of motion will be fourth order in derivatives, as explicitly
shown in \cite{aninda9}. Further even if considering the equations of
motion for graviton propagation in a general background solution or in
the AdS vacuum, these linearized equations are still fourth order. To
gain some intuition for such higher order equations, we establish an
analogy with a higher-derivative scalar field equation (in flat space)
-- following \cite{old2}. To begin, we would think of a simple massless
scalar (\ie $\Box\phi=0$) as providing the analog of the linearized
Einstein equations. Then we modify this equation with the addition of a
fourth order term to model the graviton equations produced in our
generalized gravity theory \reef{action}
 \be
 \left(\Box + \frac{a}{M^2} \Box^2\right) \phi=0\,.
 \labell{analog}
 \ee
Here we imagine $M^2$ is some high energy scale (the analog of $1/L^2$)
and $a$ is the dimensionless coupling that controls the strength of the
higher-derivative term (the analog of $\lambda_i$ and $\mu_i$). The
(flat space) propagator for this scalar can now be written as
 \be
   \frac{1}{q^2(1- a\, q^2/M^2)} = \frac{1}{q^2} -
   \frac{1}{q^2-M^2/a}\,.
 \labell{propell}
 \ee
Now the $1/q^2$ pole is associated with the regular modes which are
easily excited at low energies. The second pole $1/(q^2-M^2/a)$ is
associated with additional `physical' modes that appear at the high
energy scale. Depending on the sign of $a$, these new modes may have a
regular mass ($a<0$) or be tachyonic ($a>0$). However, the key point is
that these extra high energy modes are ghosts (for either sign of $a$)
because the overall sign of their contribution to the propagator
\reef{propell} is negative. This appearance of ghosts is a generic
feature of higher derivative equations of motion and so one must worry
that the fourth order graviton equations generically emerging from
eq.~\reef{action} indicate that these gravitational theories contain
ghosts. From a holographic perspective, this indicates that the
graviton couples to more than the usual stress tensor in the boundary
CFT. The massive ghost modes indicate that metric fluctuations also
mixes with an additional tensor operator which is non-unitary. That is,
the new operator produces states with negative norm in the CFT. Hence
from either perspective, there is a fundamental pathology with such a
theory.

However, as the analysis of quasi-topological gravity indicates
\cite{old1}, this problem can be evaded at least in the AdS vacuum.
That is, we can tune the coupling constants in eq.~\reef{action} to
special values, $\lambda^*_i$ and $\mu^*_i$, so that the linearized
graviton equations in the AdS vacuum are only second order in
derivatives. This tuning eliminates the appearance of ghosts in the
gravity theory and of non-unitary operators in the boundary CFT. In the
scalar field analogy above, specially tuned simply corresponds to
$a=0$. However, note that as we approach $a=0$ from finite values, the
mass of the ghost modes diverges. Hence in the context of our
holographic model, we can understand that the non-unitary operators are
removed from the spectrum of the boundary CFT because, as we adjust the
higher curvature coupling constants to approach the ghost-free model,
the conformal dimension of these operators diverges. Further note, that
after we have fixed the couplings to $\lambda^*_i$ and $\mu^*_i$, we
are able to calculate arbitrary $n$-point functions of the stress
tensor in the vacuum of the boundary CFT, with the usual perturbative
expansion in terms of Witten diagrams \cite{witten0}. Hence to begin,
we impose this requirement to constrain the gravitational action
\reef{action} as a tentative step towards producing a physically
interesting holographic model. Afterwards, we will examine whether RG
flows in these theories also obey a holographic c-theorem.

To identify the constraints leading to second order linearized
equations of motion for fluctuations, we proceed as follows: First, we
write the AdS$_{d+1}$ metric as\footnote{We will assume $d\geq 3$. The
$d=2$ case has been considered in \cite{sinha,paulos,aninda9,tekin}.}
 \be
ds^2=e^{2 r/\tL}\left( -dt^2+ d\vec{x}_{d-1}^2 \right)+dr^2\,.
 \labell{adsback}
 \ee
Next (using {\it Mathematica}), we consider the linearized equations of
motion around this background, including all possible metric
fluctuations $h_{\mu\nu}dx^\mu dx^\nu$ where $h_{\mu\nu}$ are allowed
to depend on all coordinates. Isolating the coefficient of
$\partial^4_r h_{12}$ in these equations, we find that this coefficient
can be set to zero with
 \be
4\lambda_1+\lambda_2+\fin\Big[3\mu_1-24\mu_2-4(d+1)\mu_3-4
d(d+1)\mu_4-(2d-1)\mu_5-3d\mu_6-d(d+1)\mu_7\Big]=0\,.
 \labell{constr1}
 \ee
Then we look at the coefficient of $\partial^4_r h_{11}$ and set this
to zero with
 \be
\lambda_1-\lambda_3
+\frac{\fin}{2}\Big[3\mu_1-12\mu_2-2d\mu_3-2(d^2+d-4)\mu_4+2\mu_5+4d
\mu_7+6d (d+1)\mu_8\Big]=0\,.
 \labell{constr2}
 \ee
Remarkably, one finds that this two constraints alone are sufficient to
eliminate all of the higher order contributions to the linearized
equations of motion! In the context of RG flows, $\fin$ will change
between the various fixed points. As a result, it is prudent to demand
that the above constraints hold for any value of $\fin$.\footnote{Note
that this condition cannot be satisfied when $d=2$, as explained in
\cite{aninda9}.} Thus we are led to the following constraints:
\begin{eqnarray}
\lambda_2&=&-4\lambda_1\,,\quad \lambda_3=\lambda_1\,,\labell{2dera}\\
\mu_7&=&\frac{1}{d(d+1)}\Big(
3\mu_1-24\mu_2-4(d+1)\mu_3-4d(d+1)\mu_4-(2d-1)\mu_5-3d\mu_6\Big)\,,
\labell{2derb}\\
\mu_8&=&\frac{1}{d(d+1)}\Big(
 -\frac{d+5}{2(d+1)}\mu_1+\frac{2(d+9)}{d+1}\mu_2+\frac{d+8}{3}\mu_3
\labell{2derc}\\
&&\qquad\qquad\qquad\qquad\quad
 +\frac{1}{3}(d^2+9d-4)\mu_4 +\frac{d-1}{d+1}\mu_5+\frac{2d}{d+1}\mu_6
\Big) \,.\nonumber
\end{eqnarray}
While eq.~\reef{2derb} follows directly from eq.~\reef{constr1}, the
constraint in eq.~\reef{2derc} comes from taking a linear combination
of the expressions appearing in both eqs.~\reef{constr1} and
\reef{constr2}. Note that the conditions \reef{2dera} on the
$\lambda_i$ yield the Gauss-Bonnet combination of curvature-squared
interactions \cite{lovel}, as expected, \ie
$(\la_1,\la_2,\la_3)\propto(1,-4,1)$.

With the constraints (\ref{2dera}--\ref{2derc}) above, we have ensured
that we have a reasonable (\ie unitary) boundary theory for the
AdS$_{d+1}$ vacua. Hence we might examine if these theories satisfy a
holographic c-theorem. So following the experience developed in the
previous section, we substitute in the RG flow geometry \reef{metric}
and examine the gravitational equation of motion proportional to
$T^t{}_t-T^r{}_r$. However, unfortunately, the resulting equation as
terms proportional to the third and fourth derivative of the conformal
factor, \ie $A'''$ and $A''''$. In order to get a simple c-theorem as
in the previous section, we can eliminate these terms by fixing
 \bea
\mu_6&=&\frac{1}{(d-1)^3}\Big(
 -2(d-3)\mu_1+8(3d-5)\mu_2+\frac23(d+1)(5d-9)\mu_3
 \labell{ccon}\\
 &&\qquad\qquad\qquad +\frac{16}3d(d+1)(d-2)\mu_4+2(d-1)^2\mu_5
\Big)\,.
 \nonumber
 \eea

Of course, with hindsight, the interpretation of this problem is
obvious. We have ensured that the non-unitary operators corresponding
to the ghost-like graviton modes have been removed from the CFT
spectrum at any fixed points. However, when the boundary theory is
perturbed away from the fixed points, the non-unitary operator come in
from infinity to `pollute' the RG flow. The solution is then also
obvious. We should demand that the linearized equations of motion for
any fluctuations around the RG flow geometry \reef{metric} are second
order in derivatives. This will ensure that the boundary QFT does not
contain any non-unitary operators along the RG flows, as well as at the
fixed points.

As before we considering general fluctuations around the RG flow metric
\reef{metric}, we examine higher order contributions to the linearized
equations of motion. If we have already imposed
eqs.~(\ref{2dera}--\ref{2derc}), the coefficients of three new terms
proportional to  $A''\, \partial_r^4 h_{11}$, $A''\,\partial_r^4
h_{12}$ and $A''\,\partial_t^4 h_{12}$ can be set to zero with
eq.~\reef{ccon} and
\begin{eqnarray}
\mu_4 &=&\frac{1}{8d}\Big(3\mu_1-12\mu_2-(d+3)\mu_3\Big)\,,\labell{2derd}\\
\mu_5 &=&-\frac{1}{d-1}\Big(12\mu_2+(d+1)\mu_3\Big)\,.\labell{2dere}
\end{eqnarray}
The seven constraints in eqs.~(\ref{2dera}--\ref{2dere}) are necessary
and sufficient to produce to two-derivative equations for metric
fluctuations around a general RG flow (\ref{metric}). This seems like
the best approach to constructing a holographic model with a physically
reasonable boundary theory in the present study of RG flows.

As noted above with eq.~\reef{ccon}, the above constraints also ensure
that, with the RG flow metric \reef{metric}, the gravitational equation
of proportional to $T^t{}_t-T^r{}_r$ takes the same simple form found
in the previous section. Hence we construct the flow function
 \bea
a_d(r)&\equiv&{\pi^{d/2}\over\Gamma\left(d/2\right)
\left(\lp A'(r)\right)^{d-1}}\, \labell{adfun2}\\
&&\times \ \left(1-{2(d-1)\over d-3}\hat\lambda L^2A'(r)^2 -
{3(d-1)\over d-5}\hat\mu L^4A'(r)^4 \right)\,. \nonumber \eea
Here, with the constraints (\ref{2dera}--\ref{2dere}), the couplings
defined in eq.~\reef{cubic8} reduce to
 \bea
 \hat\lambda&=&(d-3)(d-2)\,\lambda_1\,,\labell{confront}\\
 \hat\mu&=&\frac{(d-5)(d-3)(d-2)}{24}\left(3\,\mu_1-12\,\mu_2
 -(d-1)\,\mu_3\right)\,.
 \nonumber
 \eea
Now by construction the radial derivative of $a_d(r)$ yields
 \be
a'(r) = -\frac{\pi^{d/2}}{\Gamma\left(d/2\right)\lp^{d-1}
A'(r)^d}\left(T^t{}_t-T^r{}_r\right) \ge0\,.\labell{magic7}
 \ee
As before, we again assume the null energy condition holds to produce
the final inequality. If, as before, we denote the fixed point value of
the flow function \reef{adfun2} as
 \be
\ads\equiv a_d(r)|_{AdS}={\pi^{d/2}\tilde
L^{d-1}\over\Gamma\left(d/2\right)\lp^{d-1}} \left(1-{2(d-1)\over
d-3}\hat\lambda \fin - {3(d-1)\over d-5}\hat\mu \fin^2 \right)\,,
 \labell{astar2}
 \ee
then our higher curvature theories satisfy the holographic c-theorem
\be \left(\ads\right)_{UV}\ge \left(\ads\right)_{IR}\,. \labell{beta4}
\ee
{\it Hence we are led to conclude that demanding unitarity of the
boundary theory along the RG flows also guarantees that the theory
obeys a holographic c-theorem.}

In fact, the unitarity constraints, (\ref{2dera}--\ref{2dere}), are
more than sufficient to produce a holographic c-theorem. If we only
require that the $T^t{}_t-T^r{}_r$ equation is second order in
derivatives and that a holographic c-theorem arises, the necessary
constraints can be written as
\begin{eqnarray}
\lambda_1+\frac{d+1}{4}\lambda_2+d \lambda_3&=&0\,, \labell{conna} \\
3\mu_1+2 \mu_3+4 d\mu_4+(2d+1)\mu_5+3\mu_6+d(d+5)\mu_7 +12d^2 \mu_8 &=&0\,, \labell{connb} \\
4\mu_2+(d+1)\mu_3+4 d\mu_4+d \mu_5+\frac{d^2+1}{2}\mu_6+d(d+1)\mu_7+4 d^2 \mu_8&=&0\,.
\labell{connc}
\end{eqnarray}
Considering the curvature-squared couplings $\lambda_i$, it is easy to
see that the solution of eq.~\reef{2dera} also satisfies
eq.~\reef{conna} above. However, eq.~\reef{conna} admits a two
parameter space of solutions and so the unitary solution is only a
special case within this larger set. Clearly, analogous comments apply
for the curvature-cubed couplings $\mu_i$. However, the conclusion
seems to be that some `unphysical' models with nonunitary operators
still satisfy a holographic c-theorem. It is perhaps not too surprising
that such circumstances can arise since the RG flows only probe a small
part of the full boundary theory. What is more important is that all of
the holographic models with a unitary boundary theory are guaranteed to
satisfy a holographic c-theorem.

In fact, the above discussion is incomplete for the constraints needed
to ensure that the boundary theory is unitary. Having imposed the above
constraints, the quadratic action for gravitons in the AdS$_{d+1}$
vacua takes the form of a Fierz-Pauli action
\begin{eqnarray}\label{FP}
S&=&\frac{\tilde c}{2\lp^{d-1}}\int d^{d+1} x \sqrt{-g}\left(\frac{1}{4}
\nabla_\mu h_{\rho\lambda}\nabla^\mu h^{\rho\lambda}-\frac{1}{2}\nabla_\mu h_{\rho\lambda}
\nabla^\rho h^{\mu\lambda} +\frac{1}{2} \nabla_\mu h^{\mu\nu} \nabla_\nu h\right. \nonumber \\
&&\qquad\qquad\qquad \left. -\frac{1}{4}
\nabla_\mu h\nabla^\mu h -\frac{d(d-1)}{2\tilde L^2}(h^{\mu\nu}h_{\mu\nu}-\frac{1}{2}h^2)
+O(h^3)\right)\,.
\labell{pauli}
\end{eqnarray}
where the constant pre-factor is given by $\tilde
c=1-2\hat{\lambda}\fin -3\hat{\mu}\fin^2$. As we will see in section
\ref{other}, this coefficient controls the strength of the leading
singularity in the two-point function of the stress tensor in the
boundary CFT. In order to avoid ghost-like gravitons and to have a
unitary boundary theory, we should also impose $\tilde c>0$. However,
it is straightforward to see that this constraint is always satisfied
because of our assumption about which root of eq.~\reef{cubic9} we are
considering -- recall the discussion below eq.~\reef{cubic8}. We begin
by denoting the right-hand side of eq.~\reef{cubic9} as
$h(f)=f-\hat\lambda f^2-\hat \mu f^3$. Now we are choosing $\fin$ to be
the smallest positive root of the equation $h(f)=\alpha$. Note that
since $h(f=0)=0$ and $\alpha>0$, the function must have a positive
slope at this root, \ie $h'(f=\fin)>0$. However, recognizing that our
expression above is precisely $\tilde c=h'(f=\fin)$, we have
established that $\tilde c>0$ for this root.\footnote{One might be
concerned that we could find $h'(f=\fin)=0$ but this only occurs
outside of our chosen domain with $\lambda_i$ and $\mu_i$ `not large'
-- see a full discussion for quasi-topological gravity in \cite{old1}.}

There is a technical issue for $d=3$ and 5, which we now briefly
address. Let us focus on the case $d=3$ to be specific. Analogous
comments will apply for $d=5$ and we return to this case below. If we
examine eq.~\reef{adfun2}, it seems that our construction is singular
with $d=3$. However, this is a spurious singularity as the factor of
$1/(d-3)$ in the second term can be absorbed into our definition of
$\hat \lambda$ in eq.~\reef{confront}. More importantly, this
contribution to $a_d(r)$ is proportional to $A'(r)^2/A'(r)^{d-1}$ and
so is simply a constant in $d=3$. This is related to an obvious
ambiguity in constructing the flow function, namely, we can always add
a constant to $a_d(r)$, which, of course, does not effect the radial
evolution. While in general there is no motivation to add an extra
constant, it turns out that there is a natural choice to make here.
Quite generally, we will find that the flow function is proportional to
a Wald-type formula \cite{WaldEnt}
 \be
\frac{\partial \mathcal{L}}{\partial R^{tr}{}_{tr}}\propto A'(r)^{d-1}
a_d(r)\,.
 \labell{waldtype}
 \ee
For further discussion of this relation, we refer the reader to
sections \ref{EE1} and \ref{EE2}, as well as
refs.~\cite{sinha,aninda9}. In any event, if we want to preserve this
relation for $d=3$, we should add a constant term to the flow function
as follows
 \be
a_3(r)= \frac{\pi^2}{\lp^2 A'(r)^2}\left(1-4 \lambda_1 L^2 A'(r)^2 +3
\hat\mu L^4 A'(r)^4\right)\,.
 \labell{adfun3}
 \ee
The appearance of this constant above is related to the fact that, in
the gravity action \reef{action} with the constraint \reef{2dera}, the
curvature-squared terms are proportional to the four-dimensional Euler
density \reef{GBterm}. The latter does not effect the gravitational
equations of motion since the bulk theory is four-dimensional with
$d=3$ but it still contributes to black hole entropy
\cite{ted2}.\footnote{The full story must be more involved since it
seems that with a large coefficient, this term would lead to violations
of the second law in black hole mergers. Hence it seems that if this
topological term were to appear in the action of a complete theory of
quantum gravity, the corresponding dimensionless coupling must be
restricted to be relatively small.} Now returning to $d=5$, analogous
comments apply but it is now the contribution proportional to
$A'(r)^4/A'(r)^{d-1}$ which is a constant. We again fix this
contribution through eq.~\reef{waldtype} to produce
 \be
a_5(r) = \frac{\pi^2}{\lp^4 A'(r)^4}\left(1-4\hat\lambda L^2
A'(r)^2-3\left(3\mu_1-12\mu_2-4\mu_3\right)L^4 A'(r)^4\right) \,.
 \labell{adfun5}
 \ee

Let us examine the interactions that result in our toy model
\reef{action} after the unitarity constraints
(\ref{2dera}--\ref{2dere}) are imposed. First as already noted above,
the constraints \reef{2dera} on the $\lambda_i$ couplings require that
the curvature-squared interaction takes precisely the form of that
appearing in Gauss-Bonnet gravity \reef{GBterm}. Hence these
interactions make two-derivative contributions to the equations of
motion in a general background for $d\ge4$ \cite{lovel}. As also
discussed above, for $d=3$, this term is a topological invariant in the
four-dimensional bulk and so does not contribute to the equations of
motion. However, it still plays a role in fixing our normalization for
$a_3(r)$, as given in eq.~\ref{adfun3}. Note that for $d=2$, this
interaction simply vanishes because of a Schouten identity for
curvatures in three or fewer dimensions. One can understand this
heuristically as arising because if one attempts to evaluate
 \be
\veps^{\a1 \a2 \a3\a4}\,\veps_{\b1\b2\b3\b4}\,R_{\a1\a2}{}^{\b1\b2}
\,R_{\a3\a4}{}^{\b3\b4}
 \labell{shout}
 \ee
in less than four dimensions, the result must vanish because the
indices do not run over enough values.

Turning to curvature-cubed interactions where the eight $\mu_i$
couplings are constrained by the five equations
(\ref{2derb}--\ref{2dere}). Hence in general, one expects a
three-parameter family of unitary $R^3$ interactions, which we will
describe in terms of a basis of three independent interactions. As the
first of these, one can readily verify that the cubic Lovelock
interaction satisfies these constraints. This interaction is
proportional to the six-dimensional Euler density $\X_6$ with
 \be
\m1=-8\,,\ \m2=4\,,\ \m3=-24\,,\ \m4=3\,,\ \m5=24\,,\ \m6=16\,,\
\m7=-12\,,\ \m8=1\,.
 \labell{euler6}
 \ee
In analogy to the comments about the Gauss-Bonnet interactions, this
particular curvature-cubed interaction will only contribute
two-derivative terms to the equations of motion in a general background
for $d\ge6$ \cite{lovel}. For $d=5$, it does not contribute to the
equations of motion but still plays a role in determining $a_5(r)$, as
given in eq.~\ref{adfun5}. For $d\le4$, this term simply vanishes and
so should not be counted as one of the basis interactions.

As a second basis interaction, we can take the quasi-topological term
$\Z_{d+1}$ given in eq.~\reef{result}, which also satisfies the
constraints (\ref{2derb}--\ref{2dere}). While this term only makes two
derivative contributions to the gravitational equations in any RG flow
geometry, we should recall that fourth order terms can appear in other
backgrounds \cite{old1,old2}. Further, we should note that this term
was only constructed for $d=4$ and $d\ge6$ and so it cannot be counted
amongst the basis interactions in $d\le3$ or $d=5$.

There are, in fact, two other candidates for the third basis
interaction both of which are constructed from Weyl
tensor:\footnote{Recall the definition of Weyl tensor in  a
($d$+1)-dimensional spacetime is
 \be
W_{abcd}=R_{abcd}-{2\over d-1} \left( g_{a[c}\,R_{d]b}-g_{b[c}\,R_{d]a}
 \right)+{2\over d(d-1)}\,R\,g_{a[c}\,g_{d]b}
 \labell{while}
 \ee
using the standard notation: $X_{[ab]}=\frac{1}{2}\left( X_{ab} -
X_{ba}\right)$.}
 \be
\W_1=W_{a\,\,b}^{\,\,c\,\,\,d}\, W_{c\,\,d}^{\,\,e\,\,\,f}\,
W_{e\,\,f}^{\,\,a\,\,\,b}\,,\qquad  \W_2=W_{ab}^{\,\,\,\,\,\,cd}\,
W_{cd}^{\,\,\,\,\,\,ef}\, W_{ef}^{\,\,\,\,\,\, ab}\,.
 \label{W3x}
 \ee
In fact, these terms do not contribute to the linearized equations of
motion around the RG flow geometry \reef{metric} at all, which can be
deduced as follows. These backgrounds are conformally flat and since
these terms are cubic in the Weyl tensor, the contribution to the
quadratic action \reef{pauli} for the graviton fluctuations must be
proportional to at least one power of the Weyl tensor. Therefore there
can be no contribution to the linearized equations of motion.
Similarly, these terms do not contribute to the background equations
which determine $A(r)$ for a particular RG flow and so the
corresponding coupling constant would not appear in $\hat \mu$ in
determining the curvature scale of the fixed points \reef{cubic9} or in
the flow function \reef{adfun2}. However, we emphasize that these
interactions \reef{W3x} would effect other properties of the boundary
QFT. For example, they would contribute in a calculation of the
three-point function of the stress tensor.

Now it may seem that we have an overabundance of basis interactions,
since we have enumerated four possible unitary interactions above but
our initial count of the constraints indicated that there should only
be a three parameter family of such interactions. However, as noted in
\cite{chile}, the interactions listed above are not all independent for
$d\ge6$. In particular, in this case, we have the relation
 \be
 \Z_{d+1}=\W_1+\frac{3d^2-9d+4}{8(2d-1)(d-3)(d-4)}\left(\X_6+8\W_1
 -4\W_2\right)\,.
 \labell{redef}
 \ee
Hence we can use any three of the above interactions as our basis for
the curvature-cubed interactions in our holographic model with unitary
RG flows when $d\ge6$. For $d=5$, $Z_6$ is not defined and so our basis
would be $\X_6$, $\W_1$ and $\W_2$. For $d<5$, Schouten identities
reduce the number of number of possible interactions with $\X_6=0$ and
$\W_1=\W_2$ \cite{aninda9}. Hence for $d=4$, we would have a two
parameter family of interactions with $\Z_5$ and $\W_1$. For $d=3$,
$\Z_4$ is also not defined and so we are reduced to a one parameter
family with only $\W_1$ \cite{aninda9}.

To close this section, we re-iterate that having found that $\ads$
satisfies a c-theorem \reef{beta4} for our generalized holographic
models, one must again ask what this quantity corresponds to in the
boundary CFT. Motivated by Cardy's conjecture \reef{beta2} and our
results in the previous section, we first compare $\ads$ to the central
charge $A$ for even $d$. With the approach of \cite{adam}, we again
find a precise match
\be \ads =  A \quad{\rm for\ even\ }d\,. \labell{evend2} \ee
Hence again, we have found evidence to support Cardy's conjecture with
this broad class of holographic CFT's. However, we are again left
without an interpretation of $\ads$ for odd $d$.

\section{$\ads$ and Entanglement Entropy -- Take One} \label{EE1}

Above, we have identified a quantity $\ads$ in eq.~\reef{astar2} for a
broad class of holographic models which varies monotonically in RG
flows. For even $d$, we have shown this quantity equals the coefficient
of the A-type trace anomaly. However, a broader definition of $\ads$ is
required to interpret our results when the boundary theory has a odd
number of spacetime dimensions. We address this question here and in
the next section, where we show that $\ads$ emerges in a certain
calculation of entanglement entropy. Note that our discussion here and
in the next section does not make use of the usual holographic
calculation of entanglement entropy \cite{taka1,taka2}. The latter only
applies for Einstein gravity while here our bulk theory involves higher
curvature interactions. Of course, if we eliminate these additional
terms, our results for the entanglement entropy reduce to those
calculated with the standard approach \cite{taka1,taka2}.

We begin with the following observation: Recall quasi-topological
gravity theory \reef{ActD} for which we considered holographic
c-theorems in section \ref{one}. Black hole solutions and
thermodynamics were studied in some detail for this theory in
\cite{old1}. The horizon entropy for any of the (static) black hole
solutions found there is given by the following:
 \beq
S = \frac{2\pi}{\lp^{d-1}}\,\left(1+2\frac{d-1}{d-3}\lambda\,k
\frac{L^2}{r_h^2} -\frac{3(d-1)}{d-5}\mu\,k^2 \frac{L^4}{r_h^4}
\right)\,\oint d^{d-1}x\sqrt{h(r_h)}\,.\labell{R3Scruv}
 \eeq
where $r_h$ is the horizon radius. The final factor yields the `area'
of the horizon with $h_{ab}(r_h)$ being the induced metric on a spatial
slice of the horizon. Also $k$ is an integer with values $+1$, 0 or
$-1$ for the horizon geometry being spherical (\ie $S^{d-1}$), planar
(\ie $R^{d-1}$) or hyperbolic (\ie $H^{d-1}$), respectively. From this
result \reef{R3Scruv}, we see that with a hyperbolic horizon if we set
the horizon radius to match the AdS curvature scale, \ie $r_h = \tilde
L=L/ \fin^{1/2}$, then the factor in brackets above becomes
$1-\frac{2(d-1)}{d-3}\lambda\fin -\frac{3(d-1)}{d-5}\mu\fin^2$. From
eq.~\reef{astar}, we recognize that this is precisely the same factor
appearing in the fixed point value of our flow function. Hence, with
these choices, we find that the entropy density can be written as:
 \bea
S&=&\frac{2\pi}{\pi^{d/2}}\,\frac{\Gamma(d/2)}{\tilde L^{d-1}} \,\ads\
\int d^{d-1}x\sqrt{h(r_h)}
 \labell{prelim}\\
&=& (4\pi)^{d/2}\,\Gamma(d/2)\,T^{d-1}\, \ads\,\int
d^{d-1}x\sqrt{h(r_h)}
 \nonumber
 \eea
where $T=1/(2\pi \tilde L)$ is the Hawking temperature of the horizon
-- the latter is fixed by the choice $r_h=\tilde L$. The interpretation
in terms of the dual CFT is that eq.~\reef{prelim} gives the entropy of
a thermal bath at temperature $T$ in a background geometry $R\times
H^{d-1}$, where, as we will see, the curvature scale of the hyperbolic
geometry is precisely $1/\tilde L$.
\FIGURE[t]{
\includegraphics[width=0.8\textwidth]{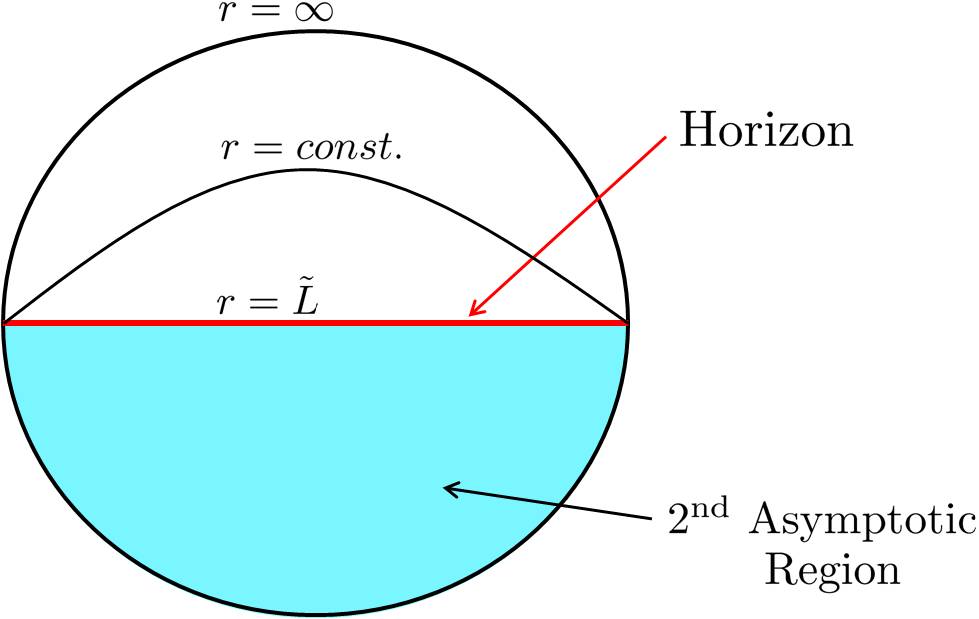}
\caption{A slice of constant $t$ through the AdS$_{d+1}$ metric in
eq.~\reef{adsmetric1}.} \label{picture}}

Note that the choice of horizon radius above also fixes the mass
parameter in the black hole metric to be zero. That is, the solution
actually reduces to a hyperbolic foliation of pure AdS$_{d+1}$, \ie
 \be
ds^2={dr^2\over \left({r^2\over\tilde L^2}-1\right)}
-\left({r^2\over\tilde L^2}-1\right)\,dt^2 + r^2\,d\Sigma_{d-1}^2
 \labell{adsmetric1}
 \ee
where $d\Sigma_{d-1}^2$ denotes the line element on $H^{d-1}$ with unit
curvature. There is a puzzle here as we have just said the bulk
geometry corresponds to the AdS vacuum and so we would expect that the
boundary theory is also in its vacuum state. From this perspective, one
must ask why should there be a nonvanishing entropy density at all? The
answer to the latter question can be found in considering the constant
time slice of eq.~\reef{adsmetric1} shown in figure \ref{picture}. This
constant $t$ slice is similar to the Einstein-Rosen bridge in a
Schwarzschild black hole \cite{HE}. The (spatial section of the) full
AdS boundary has the topology $S^{d-1}$. However, as illustrated in the
figure, with the hyperbolic foliation \reef{adsmetric1}, only half of
this boundary is reached in the limit $r\rightarrow\infty$. The other
half is reached from the second asymptotic region `behind the horizon'
at $r_h=\tilde L$. Hence a natural interpretation of the entropy is
that it corresponds to the entanglement entropy\footnote{Related ideas
were discussed in \cite{roberto,mvr}.} of the CFT (in its vacuum state)
between these two halves of the $S^{d-1}$.

Of course, to evaluate the entropy in eq.~\reef{prelim}, we must
perform the integral over the horizon. It is no surprise that the
latter yields an infinite result since, as illustrated in figure
\ref{picture}, the hyperbolic horizon extends out to the AdS boundary.
Let us write the induced metric on the bifurcation surface of the event
horizon as
 \be
r_h^2\,d\Sigma_{d-1}^2=\tL^2\,\left(\frac{d\rho^2}{1+\rho^2}
+\rho^2\,d\Omega^2_{d-2}\right)
 \labell{horizon}
 \ee
where $d\Omega^2_{d-2}$ denotes the line element on a unit
$(d-2)$-sphere.\footnote{It is useful to note that this spatial metric
on the horizon precisely matches with the spatial metric in the
boundary theory.} Integrating over the horizon out to some maximum
radius $\rho_{max}$ yields
 \be
S=\frac{2\pi}{\pi^{d/2}}\Gamma(d/2)\,
\ads\,\Omega_{d-2}\int_0^{\rho_{max}}
\frac{\rho^{d-2}\,d\rho}{\sqrt{1+\rho^2}}\,,
 \labell{stotal1}
 \ee
where $\Omega_{d-2}=2\pi^{(d-1)/2}/\Gamma((d-1)/2)$ is the area of a
unit $(d-2)$-sphere.

As already noted, the bifurcation surface of the hyperbolic horizon is
a surface that stretches across the entire AdS geometry and, so in
particular, extends out to the boundary. Hence we can interprete the
maximum radius $\rho_{max}$ above in terms of a UV cut-off in the
boundary theory. Using the standard UV/IR conversion, we have a
short-distance cut-off: $\delta=\tL/\rho_{max}$. Then we observe that
the leading contribution arising from eq.~\reef{stotal1} can be written
as
 \be
S\simeq \frac{2\pi}{\pi^{d/2}}\frac{\Gamma(d/2)}{d-2}\,\ads\,
\frac{{\mathcal A}_{d-2}}{\delta^{d-2}}+\cdots\,,
 \labell{stotal2}
 \ee
where ${\mathcal A}_{d-2}=\Omega_{d-2}\tL^{d-2}$ is the `area' of the
equator dividing the two halves of the $S^{d-1}$ in the boundary
theory. This leading divergence takes precisely the form expected for
the `area law' contribution to the entanglement entropy in a
$d$-dimensional CFT \cite{taka2,taka3}.  Note that the hyperbolic
geometry of the horizon was essential to ensure the leading power was
$1/\delta^{d-2}$ here despite the area integral being
($d\!-\!1$)-dimensional in eq.~\reef{stotal1}. This divergent
contribution to the entanglement entropy is not universal -- \eg see
\cite{taka2,taka3}. However, a universal contribution can be extracted
from the subleading terms. The form of the universal contribution to
the entanglement entropy depends on whether $d$ is odd or even
\cite{taka2,taka3}:
\be S_{univ}=\left\lbrace
\begin{matrix}
(-)^{\frac{d}{2}-1}\,\, {4}
\, \ads\, \log(2\tL/\delta)&\quad&{\rm for\ even\ }d\,,\\
(-)^{\frac{d-1}{2}}\,\, {2\pi} \, \ads\ \ \ \ \ \ \ \ \ \ \ &\quad&{\rm
for\ odd\ }d\,.
\end{matrix}\right.
\labell{unis} \ee

Up to this point the discussion focussed on quasi-topological gravity
theory \reef{ActD}, for which a broad class of black hole solutions is
known \cite{old1}. However, at the end of the day, our analysis of the
entanglement entropy only makes reference to the AdS$_{d+1}$ vacuum
solution. This suggests that our result should extend to a wider class
of gravitational theories, as we will now show in the following.

In particular, consider any of the theories considered in section
\ref{two} and present the AdS$_{d+1}$ vacuum in the hyperbolic
foliation of eq.~\reef{adsmetric1}. Treating the latter as a black
hole, the horizon entropy can be calculated using Wald's entropy
formula \cite{WaldEnt}
 \beq
S = -2 \pi \int_\mt{horizon} d^{d-1}x\sqrt{h}\
\frac{\partial{\mathcal{L}}}{\partial R^{a b}{}_{c d}}\,\hat{\veps}^{a
b}\,\hat{\veps}_{c d}\,,
 \labell{Waldformula}
 \eeq
which applies very generally for any (covariant) theory of gravity --
this was one of two approaches considered in \cite{old1} to derive
eq.~\reef{R3Scruv}. Note that above, $\mathcal{L}$ denotes the
gravitational Lagrangian and $\hat{\veps}_{a b}$ is the binormal to the
horizon. Of course, with the given metric \reef{adsmetric1}, the
integrand in eq.~\reef{Waldformula} is constant across the horizon and
so the total entropy diverges as described above. However, we use this
this constancy to rewrite eq.~\reef{Waldformula} in a way that
facilitates comparisons with eq.~\reef{prelim}
 \beq
S = -2 \pi\, \left.\frac{\partial{\mathcal{L}}}{\partial R^{a b}{}_{c
d}}\,\hat{\veps}^{a b}\,\hat{\veps}_{c d} 
\ \int d^{d-1}x\sqrt{h}\ \right|_\mt{horizon}\,,
 \labell{Waldformula2}
 \eeq
Now in section \ref{two}, we observed in eq.~\reef{waldtype} that at
any point in the RG flow geometries that the pre-factor appearing in
the above expression. To be precise, one finds
 \be
 \frac{\partial \mathcal{L}}{\partial
R^{tr}{}_{tr}}=2\frac{\Gamma(d/2)}{\pi^{d/2}}\, A'(r)^{d-1} a_d(r)\,.
 \labell{waldtype2}
 \ee
Hence if we consider an AdS$_{d+1}$ solution dual to one of the RG
fixed points, the expression above can be considered at the hyperbolic
horizon and the pre-factor in eq.~\reef{Waldformula2} becomes precisely
that appearing in eq.~\reef{prelim}. Hence all of previous discussion,
including the precise numerical factors, extends to any of the theories
in section \ref{two}. In particular then, we have the same
identification of $\ads$ with a universal coefficient in the
entanglement entropy for these theories as well.

To recap our result here: We interpret the horizon entropy of the
hyperbolic foliation of AdS$_{d+1}$ as the entanglement entropy of the
boundary CFT (in its vacuum state) between the two halves of the
$S^{d-1}$. With this framework, we are then able to identify the
coefficient $\ads$ in terms of universal contribution to the
entanglement entropy \reef{unis}. In particular though, this
identification can be applied for both {\it odd} and even $d$. Further,
while the analysis was originally made for quasi-topological gravity,
it extends to any of the gravitational theories considered in section
\ref{two}.

\section{$\ads$ and Entanglement Entropy -- Take Two} \label{EE2}

In the previous section, we have argued that $\ads$, the quantity which
appeared in the holographic c-theorem in sections \ref{one} and
\ref{two}, can be identified with a coefficient in a particular
calculation of entanglement entropy in the dual CFT. The derivation
there relied on relating the entanglement entropy for the vacuum in
$R\times S^{d-1}$ to the thermal entropy of a heat bath in $R\times
H^{d-1}$. While this discussion can be made more precise
\cite{casini9}, we would like to frame the same identification within a
more conventional calculation of entanglement entropy in this section.

Generally entanglement entropy arises as follows \cite{cardy0,taka3}:
One divides a given system into two parts, say, A and B, and integrates
out the degrees of freedom in one subsystem, B. The remaining degrees
of freedom in A are described by a density matrix $\rho_\mt{A}$. The
entanglement entropy is then simply the von Neumann entropy of this
density matrix, \ie $S=-Tr\left[\rho_\mt{A}\,\log\rho_\mt{A}\right]$.
In field theories, the system is typically subdivided by introducing a
boundary $\Sigma$ which separates the space (\ie a constant time slice)
into two regions, as shown in figure \ref{pictx}.
\FIGURE[t]{
\includegraphics[width=0.8\textwidth]{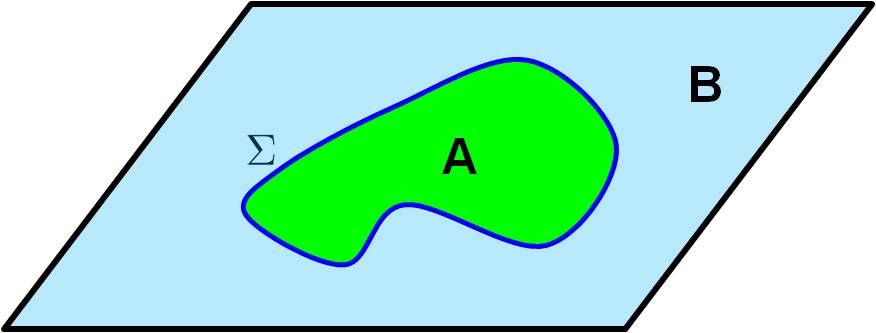}
\caption{The division of a particular time slice into two regions $A$
and $B$.} \label{pictx}}

A standard approach to calculating the entanglement entropy is to apply
the replica trick \cite{cardy0}. Since the operator $\log\rho_\mt{A}$
often lacks a clear definition, this construction begins by considering
(integer) powers of the density matrix, which may be defined as
 \be
Tr\left[\,\rho_\mt{A}^n\,\right]=Z_n/(Z_1)^n\,.
 \labell{replica1}
 \ee
Here $Z_n$ is the partition function of $n$ copies of the equivalent
system satisfying the constraint that their degrees of freedom are
identified on the region B. That is, one calculates the partition
function $Z_n$ on an $n$-fold cover of the Euclidean\footnote{An
implicit initial step in this calculation was to Wick rotate the time
coordinate to $\te=it$.} background geometry where a cut is introduced
throughout region B at $\te=0$. At the cut, the copy $n$ of the fields
is connected to copy $n$+1 when approaching from $\te\to0^-$ and to
copy $n$--1 when approaching from $\te\to0^+$. The factors of $Z_1$
appear in the denominator of eq.~\reef{replica1} to ensure that the
density matrix is properly normalized with
$Tr\left[\rho_\mt{A}\right]=1$. Assuming that the expressions
calculated in eq.~\reef{replica1} can be extended to real $n$, the
entanglement entropy can then be calculated as
  \be
S=-\lim_{n\to1}\frac{\partial\ }{\partial
n}Tr\left[\rho_\mt{A}^n\right] =-\lim_{n\to1}\left(\frac{\partial\
}{\partial n}
 -1\right)\,\log Z_n\,.
  \labell{ees1}
  \ee

Hence the calculation of the entanglement entropy can be summarized
with the following schematic outline:
\begin{enumerate}
\item Construct the $n$-fold cover described above.
\item Calculate the partition function $Z_n$ on this $n$-fold
    cover.
\item Analytically continue $Z_n$ to real $n$.
\item Evaluate the entanglement entropy using eq.~\reef{ees1}.
\end{enumerate}
While the above approach is customary in the condensed matter
literature \cite{cardy0}, a common modification is prevalent in the
high energy physics literature which gives these calculations a
geometric character \cite{callan}. Hence, we denote this the `geometric
approach' in the following. Essentially one interchanges the steps 2
and 3 above and proceeds with the following calculation:
\begin{enumerate}
\item [1.] Construct the $n$-fold cover described above.
\item [2$\,'$.] Analytically continue this background geometry to
    real $n$.
\item [3$\,'$.] Calculate the partition function $Z_n$ on the
    analytically continued cover.
\item [4.] Evaluate the entanglement entropy using eq.~\reef{ees1}.
\end{enumerate}

From this geometric perspective, the boundary $\Sigma$ is codimension
two surface where the $n$-fold cover introduces an angular excess of
$2\pi(n-1)$. The analytic continuation in step 2$\,'$ assumes that we
can define the corresponding geometry with an arbitrary angular (or
deficit) excess at this surface. Given the limit $n\to1$ in
eq.~\reef{ees1}, we need only consider values of $n$  where
$n=1-\epsilon$ with $\epsilon\ll 1$. Hence we can think that the
entanglement entropy measures the response of the field theory to an
infinitesimal conical defect at $\Sigma$.

However, we must emphasize that we are only fully justified in using
this revised calculation if there is a rotational symmetry around the
surface $\Sigma$. Without such a symmetry, the analytic continuation of
the geometry in step 2$\,'$ is not well defined.

The preceding discussion is general and so now we turn to the case of
interest which appeared in section \ref{EE1}. There we have the
$d$-dimensional boundary CFT on $R\times S^{d-1}$. As above, we Wick
rotate to Euclidean time, $\te=it$, so that the background metric
becomes
 \be
ds^2=d\te^2+\tL^2\,d\Omega_{d-1}^2\,.
 \labell{bmetric1}
 \ee
At $\te=0$, we divide the sphere into two equal halves along the
equator of $S^{d-1}$, which is then the surface $\Sigma$ in the
nomenclature introduced above. Now we would like to calculate the
entanglement entropy with the geometric approach. However, our
background geometry does not have the desired rotational symmetry
around $\Sigma$ and so as it stands this calculation would be
ill-defined. Now, we note that the field theory of interest is a CFT
and so it is possible to use the conformal symmetry to transform to a
geometry with the desired symmetry. In particular, we make the
coordinate transformation
 \be
\te=\tL\,\log\left(\frac{1-\cos\theta}{\sin\theta}\right)\,,
 \labell{transform}
 \ee
with which the metric \reef{bmetric1} becomes
 \be
ds^2=\frac{d\tilde{s}^2}{\sin^2\theta}\quad{\rm with} \quad
d\tilde{s}^2=\tL^2\left(d\theta^2+\sin^2\theta\,d\Omega_{d-1}^2 \right)
\,.
 \labell{newbmetric1}
 \ee
Hence with a conformal transformation, the background geometry has
become a $d$-dimensional sphere of radius $\tL$. Note that with the
transformation \reef{transform}, the surface $\te=0$ corresponds to
$\theta=\pi/2$ and the Weyl factor in eq.~\reef{newbmetric1} is simply
one there, \ie the conformal transformation leaves the geometry on the
$\te=0$ surface unchanged. Further $\te\to\pm\infty$ corresponds to
$\theta=0$ and $\pi$. That is, the spherical boundaries at
$\te\to\pm\infty$ in the original background geometry have been
compactified to the points at the poles of the $d$-sphere. Implicitly,
our compactification also inserts the identity operator at the poles so
that there is nothing exceptional about these two positions in the
smooth $d$-sphere geometry. These insertions can be viewed from another
perspective: Implicit in our discussion here is that the boundary field
theory is in its vacuum state. From this point of view, we can ensure
that the CFT is in its vacuum state on the $\te=0$ surface by imposing
vacuum boundary conditions at $|\te|\to\infty$ in the $R\times S^{d-1}$
background. Upon compactifying the geometry to $S^d$, inserting
identity operators at the poles simply conforms to this choice of
vacuum boundary conditions.

Of course, the $d$-sphere geometry has the desired rotational symmetry
and so we are free to consider the geometric approach for calculating
the entanglement entropy. At this point, we have set the stage to
evaluate the entanglement entropy purely from the perspective of any
conformal field theory (without referring to holography) and we turn to
a non-holographic calculation in section \ref{EEcft}. Here we continue
towards realizing this calculation in the framework of the AdS/CFT
correspondence. However, after describing the holographic calculation,
we will make a brief diversion to collect a number of apparently
unrelated results. With these, we can easily compare the results for
the entanglement entropy found here to those in the previous section.

At the outset, let us generalize the discussions. In particular, in the
following we will allow the bulk gravitational theory to be described
by any general covariant action:
 \be
I=\int d^{d+1} x \sqrt{-g}\, {\cal L}(g^{ab},R^{ab}{}_{cd}, \nabla_e
R^{ab}{}_{cd}, \dots,matter)+ boundary\ terms\,.
 \labell{covariantM}
 \ee
For the purposes of comparison, we must remind the reader that the
discussion of section \ref{EE1} entailed a Minkowski signature both in
the bulk and the boundary theory, as in eq.~\reef{covariantM}. However,
for the above calculation in the boundary theory, we have Wick rotated
to a Euclidean signature and it follows that the dual gravity
calculations will also be in Euclidean signature. Hence we wish to
distinguish the gravity action after Wick rotation as
 \be
I_\mt{E}=\int d^{d+1} x \sqrt{g}\, {\cal
L}_\mt{E}(g^{ab},R^{ab}{}_{cd}, \nabla_e R^{ab}{}_{cd}, \dots,matter)+
boundary\ terms\,.
 \labell{covariantE}
 \ee
We do so because there is a delicate issue of signs. Essentially, the
Wick rotation introduces an extra minus sign in the action so that we
have $\cL_\mt{E}=-\L$. In other words, if one evaluates the Lagrangian
density on the AdS$_{d+1}$ vacuum (as we will do later on), then one
finds $\cL<0$ in Minkowski signature while $\cL_\mt{E}>0$ in Euclidean
signature.

Now within the AdS/CFT framework, the calculation of the entanglement
entropy described for the boundary CFT above is translated to a
gravitational calculation. We begin by considering the  AdS$_{d+1}$
vacuum in Euclidean signature with metric
 \be
ds^2={dr^2\over \left({r^2\over\tilde L^2}+1\right)}
 + r^2\,d\Omega_d^2\,.
 \labell{adsvac1}
 \ee
Here $d\Omega_d^2$ denotes the line element on a $d$-sphere with unit
curvature. This foliation was chosen as it is naturally dual to the
boundary CFT on an $S^d$ background geometry (of radius $\tL$), as
desired above. As usual \cite{revue}, the path integral over the bulk
gravity fields is treated in the saddle-point approximation and hence
we have
 \be
Z_n=\exp\left[-I_{\mt{E},n}\right]
 \labell{saddle}
 \ee
where $I_{\mt{E},n}$ is the Euclidean action \reef{covariantE}
evaluated on-shell with the $n$-fold cover of the AdS$_{d+1}$ space,
where as above $n$ is real. In the previous construction with the
boundary CFT, the surface $\Sigma$ was a maximal $S^{d-1}$ in the
$d$-sphere. Constructing the $n$-fold cover then introduces a conical
singularity at $\Sigma$ with an angular excess of $2\pi(n-1)$. Here our
holographic dual extends this conical singularity to a bulk surface
$\Sigma_b$ covers the same maximal $S^{d-1}$ (in the spherical portion
of the foliation \reef{adsvac1}) and the radial coordinate in the
AdS$_{d+1}$ geometry.\footnote{Note that one does not necessarily have
to introduce a bulk singularity in these calculations of entanglement
entropy, as illustrated in \cite{mishanet2}.} However, without a full
understanding of string theory or quantum gravity in the AdS bulk, we
will not know how to resolve the conical curvature singularity produced
at finite $n$ and so it is not really possible to define the
saddle-point action in eq.~\reef{saddle}. Fortunately, as described
above, to calculate the entanglement entropy using the geometric
approach, we really only need to measure the response of the system to
an infinitesimal conical defect at $\Sigma$ and $\Sigma_b$ (\ie with
$n=1-\epsilon$ with $\epsilon\ll 1$). As we describe next, such an
infinitesimal angular deficit can be treated in a well-defined way
following the approach of \cite{furso}.

\subsection{Conical defect in AdS$_{d+1}$}

Motivated by calculations of black hole entropy, ref.~\cite{furso}
developed a description of Riemannian geometry in the presence of a
conical defect. Their prescription involved replacing the singular
geometry with a `regulator' geometry, which smoothed out the region
around the conical singularity, and then carefully considering the
limit in which this smooth geometry converges towards the original
singular manifold. This allows one to evaluate certain integrals of
covariant functionals constructed from the background geometry even in
the presence of the conical defect.

To establish some notation, let us denote the full geometry as
$\mathcal M$ and the singular surface as $\Sigma_b$, to be consistent
with the holographic discussion above. As above, we assume there is a
rotational symmetry about the singularity and we set the angle around
$\Sigma_b$ to run through a range $2\pi(1-\epsilon)$. Now the
discussion in \cite{furso} considered conical defects where $\epsilon$
was small but finite. It was found that in a general integral, such as
$\int_{\mathcal M} R^2$, the contribution linear in $\epsilon$ is
independent of the regulator but higher order terms depend on the
details of the smoothing geometry. That is, their construction
demonstrates that in such integrations, the only universally defined
result is that where conical singularity makes a linear contribution.
With this proviso, a key result of \cite{furso} is that the curvature
of $\mathcal M$ can be represented as
 \be
{}^{(\epsilon)} R^{ab}{}_{cd}=R^{ab}{}_{cd}+2\pi\epsilon\,
\sum_{i,j=1}^2\left( [n^i]^a [n^i]_c\,[n^j]^b [n^j]_d -[n^i]^a
[n^i]_d\,[n^j]^b [n^j]_c\right)\, \delta_{\Sigma_b}
 \labell{reef}
 \ee
where $R^{ab}{}_{cd}$ is the smooth contribution away from $\Sigma_b$
and $\delta_{\Sigma_b}$ is a two-dimensional $\delta$-function
satisfying $\int_{\mathcal M} 
f(x)\,\delta_{\Sigma_b} = \int_{\Sigma_b} 
f(x)$. The $[n^i]_a$ with $i=1,2$ are a pair of orthogonal unit vectors
(\ie $[n^i]_a\, [n^j]^a=\delta^{ij}$) spanning the space transverse to
$\Sigma_b$. Again, note that this formula \reef{reef} must be applied
with some care. Certainly expressions where higher powers of
$\delta_{\Sigma_b}$ appear will not be well-defined.\footnote{It was
later realized that this result also only applies in situations where
the extrinsic curvature of $\Sigma_b$ vanishes \cite{solo}.}

Now a short calculation allows us re-express eq.~\reef{reef} as follows
 \be
{}^{(\epsilon)} R^{ab}{}_{cd}=R^{ab}{}_{cd}+2\pi\epsilon\,
\tilde\veps^{ab}\,\tilde{\veps}_{cd}\, \delta_{\Sigma_b}
 \labell{reef1}
 \ee
where $\tilde\veps_{ab}\equiv[n^1]_a [n^2]_b-[n^2]_a [n^1]_b$ is the
two-dimensional volume form in the (Euclidean) space transverse to
$\Sigma_b$. This form will provide a more convenient expression for the
entanglement entropy below.

Returning to our holographic calculation, we wish to evaluate the
gravitational action for AdS$_{d+1}$ space with a conical singularity.
As illustrated in \cite{furso}, the issue of higher powers of
$\delta_{\Sigma_b}$ can be avoided with a  carefully engineered action,
such as in Lovelock gravity \cite{lovel} where the Lagrangian has a
topological origin. However, this problem certainly arises for the
general covariant action \reef{covariantE} which we are considering
here. Fortunately, we do not need to consider finite $\epsilon$ rather
we seek precisely the linear response of the system to the introduction
of the conical singularity. Combining eqs.~\reef{ees1} and
\reef{saddle}, the holographic calculation of entanglement entropy
requires that we evaluate
  \be
S=-\lim_{\epsilon\to0}\left(\frac{\partial\ }{\partial \epsilon}
 +1\right)\,I_{\mt{E},1-\epsilon}\,.
  \labell{ees2}
  \ee
With the limit $\epsilon\to0$, the (single) derivative picks out only
the linear contribution of the conical singularity in the action.

Given eq.~\reef{reef1} above, it is straightforward to show that
 \bea
I_{\mt{E},1-\epsilon}&=&(1-\epsilon)I_{\mt{E},1}+\int
d^{d+1}x\sqrt{g}\,\frac{\partial{\cL_\mt{E}}}{\partial R^{a b}{}_{c
d}}\,2\pi\epsilon\,\tilde{\veps}^{a b}\,\tilde{\veps}_{c
d}\,\delta_{\Sigma_b} +O(\epsilon^2)
 \nonumber\\
&=&(1-\epsilon)I_{\mt{E},1}+2\pi \epsilon \int_{\Sigma_b}
d^{d-1}x\sqrt{h}\,\frac{\partial{\cL_\mt{E}}}{\partial R^{a b}{}_{c
d}}\,\tilde{\veps}^{a b}\,\tilde{\veps}_{c d} +O(\epsilon^2)\,.
 \labell{react}
 \eea
Above there are two sources of $O(\epsilon)$ contributions to
$I_{\mt{E},1-\epsilon}$. First, the angle around $\Sigma_b$ runs
through the range $2\pi(1-\epsilon)$ and since we have a rotational
symmetry in this direction, the smooth contributions to the curvature
in eq.~\reef{reef1} yield the term $(1-\epsilon)I_{\mt{E},1}$. Then
there is the term linear in $\delta_\Sigma$ where each appearance of
the curvature in the Lagrangian is replaced in turn by the
$\delta$-function term on the right-hand side of eq.~\reef{reef1}. The
latter is yields the differentiation of the Lagrangian with respect to
$R^{a b}{}_{c d}$.

Now substituting the above result \reef{react} into eq.~\reef{ees2},
our AdS/CFT calculation of the entanglement entropy yields
 \beq
S = -2 \pi\, \left.\frac{\partial{\cL_\mt{E}}}{\partial R^{a b}{}_{c
d}}\,\tilde{\veps}^{a b}\,\tilde{\veps}_{c d}\ \int d^{d-1}x\sqrt{h}\
\right|_{\Sigma_b}\,.
 \labell{Waldformula3}
 \eeq
This expression is evaluated in the AdS$_{d+1}$ vacuum solution
\reef{adsvac1} (with $\epsilon=0$). Hence the pre-factor above is
constant and so we have pulled it out of the integral over $\Sigma_b$.

We have expressed this result \reef{Waldformula3} in a way that it
appears to have the same form as the formula for the Wald entropy in
eq.~\reef{Waldformula2}. Of course, the present discussion made no
reference to a black hole geometry or black hole thermodynamics. We
must also emphasize that there are subtle differences arising because
the spacetime signature differs between the two expressions. In
particular then, eq.~\reef{Waldformula2} contains the variation of the
Minkowski-signature Lagarangian $\cL$ while eq.~\reef{Waldformula2}
contains that for Euclidean-signature Lagarangian $\cL_\mt{E}$. As
described above, there is a difference of sign between these two, \ie
$\cL_\mt{E}=-\cL$. However, this minus sign is compensated by an extra
sign that arises in Wick rotating the two transverse volume forms
appearing in each expression \eg in Minkowski signature, we have
$\hat{\veps}^{a b}\,\hat{\veps}_{ab}=-2$ while $\tilde{\veps}^{a
b}\,\tilde{\veps}_{ab}=2$ in Euclidean signature. Further, one can
verify that upon analytically continuing the Euclidean AdS$_{d+1}$
metric back to Minkowski signature, the surface $\Sigma_b$ above
coincides with the bifurcation surface of the hyperbolic black hole
considered in section \ref{EE1}. Hence the expression in
eq.~\reef{Waldformula3} above and that in eq.~\reef{Waldformula2} are
evaluated on precisely the same geometry and yield precisely the same
result for the entanglement entropy.\footnote{The same analysis can be
applied in a discussion of black hole entropy and our results readily
confirm the equivalence of Wald's Noether charge method \cite{WaldEnt}
and the off-shell approach \cite{offish}, which introduces a conical
singularity in the Euclidean black hole geometry. The equivalence of
these two techniques was previously discussed in \cite{willie}.} We
emphasize again that the results here and in the previous section apply
for a completely general covariant gravity theory \reef{covariantM}. We
will need two more results to connect eq.~\reef{Waldformula3} to $\ads$
in this general context.

\subsection{AdS equations of motion}

Next we consider the equations of motion arising from the general
action \reef{covariantM}. In particular, we will organize the field
equations so as to separate the contributions coming from the
variations of the (inverse) metric $g^{ab}$ and the curvature
$R^{ab}{}_{cd}$. Hence we write the variation of the action
\reef{covariantM} as:
 \be
\delta I=\int d^{d+1} x \sqrt{-g}\,\left[-{1\over2}{\cal
L}\,g_{ab}\,\delta g^{ab}
+\frac{\delta \cal L}{\delta g^{ab}}\,\delta g^{ab}
+ \frac{\delta \cal L}{\delta R^{ab}{}_{cd}}\,\delta
R^{ab}{}_{cd}\right]
 \labell{eom}
 \ee
Now a few comments are in order here: First we stress that we are not
treating the curvature $R^{ab}{}_{cd}$ as an independent field. We are
still making the standard variations above with
 \bea
\delta R^{ab}{}_{cd}&=&g^{be}(\nabla_c \delta \Gamma^a_{\ ed}-\nabla_d
\delta \Gamma^a_{\ ec})+R^a_{\ ecd}\delta g^{be}\,,
 \labell{vary0}\\
 \delta \Gamma^a_{\
ed}&=&\frac{1}{2}g^{af}(\nabla_e \delta g_{fd}+\nabla_d\delta
g_{fe}-\nabla_f\delta g_{ed})\,,
 \labell{vary1}
 \eea
and $\delta g_{ab}=-g_{ac}\,g_{bd}\,\delta g^{cd}$. Our goal in writing
eq.~\reef{eom} is simply to organize the various contributions in
equations of motion and in particular we wish to isolate those coming
from the variations $\delta R^{ab}{}_{cd}$. In particular, the
expression ${\delta \cal L}/{\delta R^{ab}{}_{cd}}$ above in
eq.~\reef{eom} is precisely same as that appearing in Wald's entropy
formula \reef{Waldformula}. Another comment is that the notation above
hides the result of any necessary integration by parts in ${\delta \cal
L}/{\delta g^{ab}}$ and ${\delta \cal L}/{\delta R^{ab}{}_{cd}}$.
Finally, let us observe that given this approach, ${\delta \cal
L}/{\delta g^{ab}}$ collects contributions from two sources: variations
of the connections \reef{vary1} appearing in covariant derivatives of
the curvature tensor appearing in the action and variations of explicit
metrics and connections which may appear in interactions involving the
matter fields. Now it is straightforward to write out the general
gravity equations of motion as
 \be
{1\over2}{\cal L}\,g_{ab}=
\frac{\delta \cal L}{\delta g^{ab}}
+ 2\,g_{(a|d}\nabla^e\,\nabla_c\frac{\delta \cal L}{\delta
R^{b)e}{}_{cd}} - R^{e}{}_{(a|cd}\,\frac{\delta \cal L}{\delta
R^{b)e}{}_{cd}}\,.
 \labell{geneom}
 \ee

Next, we wish to consider evaluating the equations in an AdS$_{d+1}$
background which produces a number of simplifications. In particular,
this background is a maximally symmetric geometry. Further any
background tensors constructed from the geometry and/or the matter
fields must respect the same symmetries and so in particular, any
covariant derivatives of such quantities will vanish in general, \ie
$\nabla_a[\cdots]=0$. Hence the second term on the right-hand side of
eq.~\reef{geneom} must vanish in this background. Further, it is clear
that all of the contributions in ${\delta \cal L}/{\delta g^{ab}}$ will
also automatically vanish in the AdS$_{d+1}$ background. Finally,
turning to the third term on the right-hand side, we set
 \be
R^{ab}{}_{cd}=-\frac{1}{\tL^2}\left(\delta^a{}_c\,
\delta^b{}_d-\delta^a{}_d\,\delta^b{}_c\right)
 \labell{adscurvature}
 \ee
for the AdS$_{d+1}$ background. Now combining these three
simplifications, the equations of motion \reef{geneom} reduce to
 \be
\left.\frac{\delta \cal L}{\delta
R^{(a|e}{}_{cd}}\right|_{AdS}\,g_{b)c}\,\delta^e{}_d= -{\tilde
L^2\over4}\left.{\cal L}\right|_{AdS}\,g_{ab}\,.
 \labell{adseom1}
 \ee
Hence we see that, in the AdS$_{d+1}$ background, the equations of
motion balance two sets of nontrivial contributions. We stress that
despite our assumption that the background geometry corresponds to
AdS$_{d+1}$, eq.~\reef{adseom1} is still a nontrivial equation. In
particular, it determines the curvature scale for the vacuum
solution(s).

We make one further step at this point. As commented above, all of the
background tensors must respect the symmetries of the AdS$_{d+1}$
background. Therefore we must have ${\delta \cal L}/{\delta
R^{ab}{}_{cd}}\propto (\delta_a{}^c\, \delta_b{}^d -\delta_a{}^d\,
\delta_b{}^c)$ and we can re-express eq.~\reef{adseom1} as
 \be
\left.\frac{\delta \cal L}{\delta R^{ab}{}_{cd}}\right|_{AdS}= -{\tilde
L^2\over4\,d}\left.{\cal L}\right|_{AdS}\,\left(\delta_a{}^c\,
\delta_b{}^d -\delta_a{}^d\, \delta_b{}^c\right)\,.
 \labell{adseom2}
 \ee

Of course, the signature of the bulk spacetime has no effect on the
preceding analysis and hence the same result follows equally well from
the Euclidean signature action \reef{covariantE}. For completeness, we
explicitly write out the corresponding expression:
 \be
\left.\frac{\delta \cL_\mt{E}}{\delta R^{ab}{}_{cd}}\right|_{AdS}=
-{\tilde L^2\over4\,d}\left.\cL_\mt{E}\right|_{AdS}\,\left(\delta_a{}^c
\delta_b{}^d -\delta_a{}^d \delta_b{}^c\right)\,.
 \labell{adseom2E}
 \ee

\subsection{A-type trace anomaly}

Given our results in eqs.~\reef{evend} and \reef{evend2}, we anticipate
that the result for the entanglement entropy should be proportional to
the $A$ coefficient in the trace anomaly \reef{trace} for even $d$.
Here we simply point out the short cut to calculating $A$ presented in
\cite{adam}. Given a general action \reef{covariantM} for the bulk
gravity theory, $A$ can be determined by simply evaluating value the
Lagrangian in the AdS$_{d+1}$ vacuum. With the present conventions, we
have
 \be
A=-\frac{\pi^{d/2}\,\tilde L^{d+1}}{d\,\Gamma\left(d/2 \right)}
\left.{\cal L}\right|_{AdS}\,.
 \labell{adamA}
 \ee
While this result was derived for even $d$, one can confirm that same
result applies for $\ads$ in odd $d$ with quasi-topological gravity
\reef{ActD} or any of the gravitational theories considered in section
\ref{two}. Hence we will use eq.~\reef{adamA} to provide a simple
definition for $\ads$ for a general gravitational theory
\reef{covariantM} in both odd and even dimensions, \ie
 \be
\ads\equiv-\frac{\pi^{d/2}\,\tilde L^{d+1}}{d\,\Gamma\left(d/2 \right)}
\left.{\cal L}\right|_{AdS}\,.
 \labell{adamA2}
 \ee

In this case, the signature of the bulk spacetime does affect the
overall sign in this result. Hence in terms of the the Euclidean
signature Lagrangian, one finds:
 \be
\ads\equiv\frac{\pi^{d/2}\,\tilde L^{d+1}}{d\,\Gamma\left(d/2 \right)}
\left.\L_\mt{E}\right|_{AdS}\,.
 \labell{adamA2E}
 \ee

 \smallskip
We have now established our three key results,
eqs.~\reef{Waldformula3}, \reef{adseom2E} and \reef{adamA2E}. Recall
that these all apply for a completely general gravity theory
\reef{covariantE}. Combining these equations, we arrive at the
following expression for the entanglement entropy
 \be
S=\left.\frac{2\pi}{\pi^{d/2}}\, \frac{\Gamma(d/2)}{\tilde L^{d-1}}
\,\ads\ \int d^{d-1}x\sqrt{h}\ \right|_{\Sigma_b}\,.
 \labell{sfinal1}
 \ee
Again we observe that the surface $\Sigma_b$ in the Euclidean
AdS$_{d+1}$ geometry coincides with the bifurcation surface of the
hyperbolic black hole considered in section \ref{EE1} upon analytic
continuation back to Minkowski signature. Hence our result
\reef{sfinal1} here precisely matches that given in eqs.~\reef{prelim}
and \reef{stotal1}. Hence we recover the area law in eq.~\reef{stotal2}
and the universal contribution to the entanglement entropy
\be S_{univ}=\left\lbrace
\begin{matrix}
(-)^{\frac{d}{2}-1}\,\, {4}
\, \ads\, \log(2\tL/\delta)&\quad&{\rm for\ even\ }d\,,\\
(-)^{\frac{d-1}{2}}\,\, {2\pi} \, \ads\ \ \ \ \ \ \ \ \ \ \ &\quad&{\rm
for\ odd\ }d\,.
\end{matrix}\right.
\labell{unis2} \ee
Here we have established the role of $\ads$ in the holographic
entanglement entropy for a completely general bulk theory of gravity.
However, in this general context, one cannot expect that $\ads$ always
satisfies the holographic c-theorem \reef{beta3}. The latter would
still require that the gravitational couplings are constrained along
the lines discussed in section \ref{two}.

\subsection{Entanglement entropy without holography} \label{EEcft}

Here we set aside our discussion of holographic entanglement entropy
and instead consider a purely field theoretic calculation of
entanglement for the configuration considered above. For a conformal
field theory in an even number of spacetime dimensions, the universal
coefficient in the entanglement entropy can be determined through the
trace anomaly. This approach relies on the geometric approach
\cite{callan} for calculating the entanglement entropy that was
discussed at the beginning of this section. The connection to the trace
anomaly first appeared in calculations for two-dimensional CFT's
\cite{finn} -- see related ideas in \cite{cardy1x} -- and later
extended to higher dimensions \cite{taka2,solo}.  In the following, our
discussion follows closely that presented in \cite{taka2}.

As above, we begin with considering the case of a $d$-dimensional CFT
on $R\times S^{d-1}$. Dividing the sphere into two along the equator
$\Sigma$, \ie along a maximal $S^{d-2}$, we wish to determine the
universal coefficient in the entanglement entropy between the two
halves of the sphere. Recall that with the geometric approach, we must
evaluate the partition function on a background geometry with an
infinitesimal conical defect. In order to construct a symmetric
geometry where introducing such a defect is well-defined, we
conformally transform the (Euclidean) background metric \reef{bmetric1}
to a $d$-sphere metric \reef{newbmetric1}. Given the rotational
symmetry around $\Sigma$ with the latter metric, we construct ${\cal
M}_{1-\epsilon}$, the `$n$-fold cover' with $n=1-\epsilon$, by
introducing an infinitesimal conical defect at $\Sigma$. The expression
for the entanglement entropy \reef{ees1} then becomes
  \be
S=\lim_{\epsilon\to0}\left(\frac{\partial\ }{\partial \epsilon}
 +1\right)\,\,\log Z_{1-\epsilon}\,.
  \labell{ees4}
  \ee

Now we observe that there is a single scale in the particular geometry
for which we are calculating the entanglement entropy, \ie $\tL$, the
radius of curvature of the original $S^{d-1}$ in \reef{bmetric1}. To
connect the calculation to the trace anomaly, we consider shifting this
scale with \cite{taka2}\footnote{For readers following the details
here, we emphasize that these calculations are made for a background
with Euclidean signature.}
 \bea
\tL\,\frac{\partial S}{\partial
\tL}&=&\lim_{\epsilon\to0}\left(\frac{\partial\ }{\partial \epsilon}
 +1\right)\,\,\tL\,\frac{\partial\ }{\partial
\tL}\log Z_{1-\epsilon}
 \nonumber\\
&=&\lim_{\epsilon\to0}\left(\frac{\partial\ }{\partial \epsilon}
 +1\right)\,\,\int d^dx \sqrt{g}\,\langle\, T^a{}_a\,\rangle
 \labell{trace9}\\
&=&\lim_{\epsilon\to0}\left(\frac{\partial\ }{\partial \epsilon}
 +1\right)\,\,\int d^dx \sqrt{g}\,\left[ \sum B_i\,I_i -2\,(-)^{d/2}A\, E_d\right]\,.
 \nonumber
 \eea
In the last line, we have used eq.~\reef{trace} to replace the trace of
the stress tensor. Now we observe that the resulting expression has
essentially the same form as eq.~\reef{ees2}, in that, we are measuring
the linear response to $\epsilon$ of an integral of some scalar
constructed with curvatures and covariant derivatives in the background
geometry. Hence we can follow the analysis presented in the holographic
calculation and we arrive at the following analog of
eq.~\reef{Waldformula3} for the entanglement entropy
 \beq
\tL\,\frac{\partial S}{\partial \tL} = 2 \pi\, \int
d^{d-2}x\sqrt{h}\,\tilde{\veps}^{a b}\,\tilde{\veps}_{c d}\ \left[\sum
B_i\,\frac{\partial I_i}{\partial R^{a b}{}_{c d}} -2\,(-)^{d/2}A\,
\frac{\partial E_d}{\partial R^{a b}{}_{c d}}\ \right]_{\Sigma}\,.
 \labell{Waldformula9}
 \eeq

At this point, we must examine the expressions entering the trace
anomaly \reef{trace} in more detail. Recall that $E_d$ is the Euler
density in $d$ dimensions and $I_i$ are independent Weyl invariants
constructed from contractions of $d/2$ curvatures or $d/2-k$ curvatures
and $2k$ covariant derivatives. There is some ambiguity in constructing
the invariants $I_i$, however, a general construction was studied in
\cite{feffer}. In this approach, the natural building blocks of the
invariants are the Weyl tensor $W_{abcd}$, the Cotton tensor $C_{abc}$
and the Bach tensor $B_{ab}$ (as well as covariant derivatives of
these).\footnote{As the latter two tensors may be unfamiliar to some
readers, we recall that
 \bea
C_{abc}&=&\nabla_cR_{ab}-\nabla_bR_{ac}+{1\over 2(d-1)}
\left(g_{ac}\nabla_bR-g_{ab}\nabla_cR\right)\,,
 \labell{defs}\\
B_{ab}&=&P_{cd}W_a{}^c{}_b{}^d+\nabla^c\nabla_aP_{bc} -\nabla^2P_{ab}
\quad{\rm with}\ \ P_{ab}={1\over d-2}\left(R_{ab}-{R\over
2(d-1)}g_{ab}\right)\,.
 \nonumber
 \eea}
Next we observe that these basis tensors all vanish for the spherical
metric \reef{newbmetric1}, \ie $W_{abcd}=0 =C_{abc}=B_{ab}$ for a round
$S^d$.\footnote{As an aside, let us note that simple result which
follows from this construction is that the only contribution to the
trace anomaly in $d$ dimensions is $A$, \ie $\oint_{S^d}\langle\,
T^a{}_a\, \rangle= 4(-)^{(d-2)/2} A$.} Further with this construction,
all of the $I_i$ are at least quadratic in these three tensors and as a
result, $\partial I_i/\partial R^{a b}{}_{c d}$ is at least linear in
one of these tensors. Therefore we can conclude that all of the
contributions in eq.~\reef{Waldformula9} proportional to the
coefficients $B_i$ vanish in the present calculation. This leaves us
only to consider the contribution from the A-type anomaly.

Our convention for the normalization of the Euler density $E_d$ in
eq.~\reef{trace} is that on a $d$-dimensional sphere:
$\oint_{S^d}d^d\!x\sqrt{g}\, E_d =2$. With this normalization, it is
straightforward to show that\footnote{Here one can use related results
for Lovelock gravity in \cite{ted2,furso}. Recall, however, that the
present calculations are performed for a Euclidean signature.}
 \be
2 \pi\, \tilde{\veps}^{a b}\,\tilde{\veps}_{c d}\, \frac{\partial
E_d}{\partial R^{a b}{}_{c d}}=E_{d-2}\,.
 \labell{euler}
 \ee
which we then substitute into eq.~\reef{Waldformula9}. Recall that the
conical singularity lies on a maximal $S^{d-2}$ and hence integrating
over this surface, we find the following simple result:
 \be
\tL\,\frac{\partial S}{\partial \tL}=(-)^{\frac{d}{2}-1}\, {4} \, A\,.
 \labell{almost}
 \ee
If we integrate this expression with respect to $\tL$, we arrive at
 \be
S=(-)^{\frac{d}{2}-1}\, {4} \, A\,\log(\tL/\delta)\,,
 \labell{final9}
 \ee
where $\delta$ is the short-distance cut-off that we use to regulate
the calculations. Hence we see that the coefficient of the universal
term in the entanglement entropy is proportional to the central charge
$A$. Note that this result \reef{final9} does not quite match the
holographic results for even $d$ presented in eqs.~\reef{unis} and
\reef{unis2}. However, the difference can be simply regarded as a
non-universal constant term.

In general, this approach to calculating the entanglement entropy (with
even $d$) yields a result depending on all of the central charges
appearing in the trace anomaly \cite{solo}. Here we have chosen a very
specific background geometry and a specific boundary dividing this
space and we have found that the universal coefficient in entanglement
entropy \reef{final9} singles out precisely $A$, the central charge of
the Euler density in eq.\reef{trace}. This matches our holographic
result \reef{unis2} since we have also identified $\ads=A$ for even
$d$. However, eq.~\reef{final9} is a general statement about $S_{univ}$
on $S^{d-1}\times R$ with any CFT in even $d$.

Let us add a few additional remarks about this calculation: Firstly,
the general approach employed above can only be successfully applied
when the system for which we are calculating the entanglement entropy
contains a single scale. This is certainly the case for the present
case, where the scale $\tL$ characterizes both the curvature of the
background sphere and the size of the equator dividing this $S^{d-1}$
into two halves. Further, the rotational symmetry around the equator of
the $d$-sphere ensures that the extrinsic curvature of this surface
vanishes. Otherwise we should have expected that additional
`corrections' involving the extrinsic curvature should be added to the
result in eq.~\reef{Waldformula9} \cite{solo}. Also, note that in
making use of the trace anomaly in eq.~\reef{trace9}, the calculation
works with the renormalized stress tensor and so we have implicitly
discarded divergent terms that might have naturally appeared with
inverse powers of $\delta$. Hence the present calculation has
eliminated the non-universal terms that might have appeared in the
entanglement entropy, \eg the leading `area' term. Finally if this
calculation was performed in an odd number of spacetime dimensions, the
result would vanish because the trace anomaly is zero for odd $d$.
However, this is in keeping with the expectation that there is no
logarithmic contribution to the entanglement entropy for odd $d$.

\subsection{Counting degrees of freedom} \label{density}

Ref.~\cite{lenny} considered how the AdS/CFT correspondence realized
the `holographic principle' \cite{holo,bousso}. Of course, the latter
proposes that in a consistent theory of quantum gravity, the physics
within a macroscopic region of space can be described by a dual theory
living on the boundary of that region. However, central element of this
proposal is that `the boundary theory should not contain more than one
degree of freedom per Planck area.' While the first aspect of the
proposal is obviously realized in the AdS/CFT framework,
ref.~\cite{lenny} examined the latter for the specific example of the
correspondence between $N$=4 super-Yang-Mills (SYM) and Type IIb
strings in AdS$_5\times S^5$. As the boundary theory is a CFT, it
contains an infinite number of degrees of freedom unless one introduces
a short distance cut-off, $\delta$. As the SYM theory has a $U(N_c)$
gauge group, then one has roughly $N_\mt{dof}\simeq N_c^2\,
V_3/\delta^3$ in the regulated theory, where $V_3$ is the spatial
volume. In the AdS$_5$ geometry, the short distance cut-off is
associated with a regulator surface at some large radius
$R=\tL^2/\delta$. Then using the standard AdS/CFT dictionary, the
authors of \cite{lenny} show
 \be
N_\mt{dof}\simeq N_c^2\, \frac{V_3}{\delta^3}\simeq\frac{{\cal
A}_3(R)}{\lp^3}
 \labell{old}
 \ee
where ${\cal A}_3(R)$ is the `area' of (a constant $t$ slice of) the
regulator surface. This result then confirms that both of the desired
aspects of the holographic principle are realized in this example of
the AdS/CFT correspondence.

Implicitly above, we are considering a regime where the bulk string
theory is well approximated by Einstein gravity in AdS$_5$. Then we see
in eq.~\reef{old} that the Bekenstein-Hawking entropy evaluated on a
holographic screen at some large $R$ reproduces the counting of the
degrees of freedom in the regulated boundary CFT. Let us now apply this
reasoning in the more general framework with higher curvature gravity
in the bulk that we have been studying. Rather than using the
Hawking-Bekenstein entropy here, it is natural to evaluate Wald's
entropy formula \reef{Waldformula} on the holographic screen near the
AdS$_{d+1}$ boundary. Then following the reasoning above but in
reverse, we expect that this geometric result will count the number of
degrees of freedom in the (regulated) boundary theory.

Let us begin with the AdS$_{d+1}$ metric in global coordinates
 \be
ds^2={dr^2\over \left({r^2\over\tilde L^2}+1\right)}
-\left({r^2\over\tilde L^2}+1\right)\,dt^2 + r^2\,d\Omega_{d-1}^2\,,
 \labell{adsmetric2}
 \ee
where $d\Omega_{d-1}^2$ is the metric on a unit ($d$--1)-sphere.
Extracting the conformal factor $r^2/\tL^2$ from the asymptotic metric
along the CFT directions as usual, we arrive at the following metric
for the boundary theory:
 \be
ds^2=-dt^2+\tL^2\,d\Omega_{d-1}^2\,.
 \labell{bmetric2}
 \ee
Next we calculate the Wald entropy on a holographic screen near the
boundary. That is, the expression in eq.~\reef{Waldformula} is
evaluated on an $S^{d-1}$ at some large radius $R$ (and constant $t$).
Interpreting the result as the number of degrees of freedom in the
boundary theory, we have
 \beq
N_\mt{dof} = -2 \pi\, \left.\frac{\partial{\mathcal{L}}}{\partial R^{a
b}{}_{c
d}}\,\hat{\veps}^{a b}\,\hat{\veps}_{c d} 
\ \int d^{d-1}x\sqrt{h}\ \right|_{(r=R,t=constant)}\,,
 \labell{Waldformula5}
 \eeq
where the pre-factor above is again constant. Note that, just as in the
Wald expression \reef{Waldformula} for black hole entropy,
$\hat{\veps}_{a b}$ is the two-dimensional volume form in the space
transverse to the surface on which the entropy is evaluated. Given the
metric \reef{adsmetric2}, the `area' integral yields $\int
d^{d-1}x\sqrt{h}=R^{d-1}\Omega_{d-1}= (R/\tL)^{d-1} V_{d-1}$ where
$V_{d-1}$ is the volume of the corresponding surface in the boundary
CFT with eq.~\reef{bmetric2}. With the standard general arguments
\cite{lenny,joep} applied to our generalized holographic framework, we
derive the usual relation between the radius $R$ in AdS$_{d+1}$ and a
short-distance cut-off $\delta$ in the boundary theory, \ie
$R=\tL^2/\delta$. Hence eq.~\reef{Waldformula5} can be re-expressed as
 \bea
N_\mt{dof} &=& -2 \pi\,\left(\frac{\tL}{\delta}\right)^{d-1}\,V_{d-1}\
\left.\frac{\partial{\mathcal{L}}}{\partial R^{a b}{}_{c
d}}\,\hat{\veps}^{a b}\,\hat{\veps}_{c d} \
\right|_{(r=R,t=constant)}
 \nonumber\\
%
%
&=& \frac{2 \pi}{\pi^{d/2}}\Gamma(d/2)\ \ads\
\frac{V_{d-1}}{\delta^{d-1}}\,.
 \labell{Waldformula6}
 \eea
Recalling that the background geometry is simply AdS$_{d+1}$, we have
applied eqs.~\reef{adseom2} and \reef{adamA2}, as well as
$\hat{\veps}^{a b}\,\hat{\veps}_{ab}=-2$, in producing the final
expression.

Hence our expression \reef{Waldformula6} for the number of degrees of
freedom is very similar in form to that \reef{old} for the $N$=4 SYM.
In particular, the factor of $V_{d-1}/\delta^{d-1}$ counts the number
of minimum size cells in regulated boundary theory, as above. Further
$\ads$ replaces the factor of $N_c^2$ in counting the number of degrees
of freedom in each such cell. Hence the present calculation suggests
that $\ads$ provides a measure of the density of the degrees of freedom
in the boundary CFT. Since we are beginning with Wald's formula in
eq.~\reef{Waldformula5}, our result does not technically match the `one
degree of freedom per Planck area' rule but realizes it in spirit.
After all, this rule emerged from considerations of the
Bekenstein-Hawking entropy of black holes in Einstein gravity and here
we are taking the natural generalization to higher curvature theories.

\section{Other central charges} \label{other}

In sections \ref{EE1} and \ref{EE2}, we were able to identify $\ads$
with the universal coefficient in a certain calculation of entanglement
entropy. Further, for even $d$, we showed that $\ads$ precisely matches
$A$, the central charge multiplying the Euler density in the trace
anomaly \reef{trace}. One of the advantages of working with a
holographic model where the bulk gravity theory contains higher
curvature interactions is that we can distinguish between the different
central charges in the dual CFT. Hence we are also able to show that
$\ads$ does not agree with various candidate charges that have been
considered in examining possible c-theorems in higher dimensions.

One proposal for a c-theorem \cite{osborn} is to consider the
coefficient characterizing the leading singularity in the two-point
function of two stress tensors \cite{osborn,Dobrev}:
 \beq
\langle\, T_{ab}(x)\, T_{cd}(0)\, \rangle =\frac{\ct}{x^{2d}}\
{\mathcal I}_{ab,cd}(x)\,,
 \labell{twopt}
 \eeq
where
 \beq
 {\mathcal I}_{ab,cd}(x)=\frac{1}{2}
\left( I_{ac}(x)I_{bd}(x)+I_{ad}(x)I_{bc}(x)\right) -{1\over
d}\eta_{ab}\,\eta_{cd}\,,\labell{tensor0}
 \eeq
and
 \beq
I_{ab}(x)=\eta_{ab}-2\frac{x_a\,x_b}{x^2}\,.
 \labell{tensor}
 \eeq
This coefficient $\ct$ is a ``central charge" common to CFT's in any
number of dimensions, $d$. Of course, up to the normalization, $\ct$
also corresponds to a standard central charge appearing in the trace
anomaly, \eg  $\ct=4\,c$ in eq.~\reef{trace2} for $d=2$ and
$\ct=(40/\pi^4)\,c$ in eq.~\reef{trace4} for $d=4$. As such, various
counter-examples to this proposed c-theorem exist for four dimensions
\cite{count1,anselmi} and so it cannot hold in complete generality. In
any event, it is straightforward to calculate the two-point function
\reef{twopt} in the dual gravity theory, \eg see \cite{old2,holoGB},
and using the notation of section \ref{two}, we find
 \be
\ct= \frac{d+1}{d-1}\frac{\Gamma[d+1]}{\pi^{d/2}\Gamma[d/2]}\,
\frac{\tL^{d-1}}{\lp^{d-1}}(1-2\fin\hat\lambda-3\fin^2\hat\mu)\,.
 \labell{ctfinal}
 \ee
Here it is clear that $\ads$ and $\ct$ are not related (in any simple
way) in our holographic models.

In considering extensions of the c-theorem to higher dimensions,
another proposal \cite{hot0,hot1} was to consider the leading
coefficient of the free energy density at finite temperature. Here we
define this `central charge' $C_S$ in terms of the thermal entropy
density using the fact that $s/T^{d-1}$ provides a dimensionless ratio
characterizing a CFT in $d$ dimensions. We adopt the normalization of
\cite{hotx0} in defining $C_S$ as
 \beq
C_S\equiv \frac{d+1}{d-1}\left(\frac{d}{2\pi^{3/2}}\right)^d
\frac{\Gamma\left((d+1)/2\right)}{\sqrt{\pi}}\ \frac{s}{T^{d-1}}\,.
 \labell{centrals0}
 \eeq
With this normalization, $C_S$ is again related to the usual central
charge in two dimensions by a simple numerical factor, \ie $C_S=4\,c$.
However, there is no simple relation between $C_S$ and any other
central charges characterizing a strongly coupled CFT in higher
dimensions, \eg see \cite{hotx0,hotx1}. We have $C_S/\ct=1$ for
strongly coupled $N=4$ SYM in four dimensions while at weak coupling,
the SYM theory yields $C_S/\ct=4/3$ \cite{hotx1} -- see also
\cite{holoGB} for further discussion. The strong coupling result
$C_S/\ct=1$ applies for any holographic CFT dual to Einstein gravity.
Of course, it is derived by studying the thermal behaviour of planar
AdS$_{d+1}$ black holes in the gravity theory. Unfortunately we do not
have solutions describing these black holes for the general theories
considered in section \ref{two}. Hence we focus our attention on
quasi-topological gravity studied in section \ref{one} and in this
case, the necessary black holes and their thermodynamic properties were
studied in \cite{old1}. Given the normalization in
eq.~\reef{centrals0}, we find
 \be
C_S= \frac{d+1}{d-1}\frac{\Gamma[d+1]}{\pi^{d/2}\Gamma[d/2]}\,
\frac{\tL^{d-1}}{\lp^{d-1}}\left(\frac{\fin}{\alpha}\right)^{d-1}
 \labell{csfinal}
 \ee
where $\alpha$ is the parameter appearing in the cosmological term in
the action \reef{ActD}. Using eq.~\reef{cubic}, we can write
 \be
\frac{\fin}{\alpha}=\frac{1}{1-\lambda\fin-\mu\fin^2}\,.
 \labell{ratio1}
 \ee
Given the expression \reef{csfinal} for $C_S$, it is again clear that
$\ads$ is not related to $C_S$ (in any simple way) for our holographic
models.

One may further ask whether $\ads$ can be determined by some $n$-point
function of the stress tensor, in particular for $d$ odd. Implicitly,
such a result already appears with $d$ even. There we have shown
$\ads=A$ and we know that $A$ can be identified as a particular
coefficient in a ($d/2$+1)-point function \cite{arkady}.\footnote{The
other the central charges $B_i$ in the trace anomaly are similarly
determined by certain two-, three-, $\ldots$, $d/2$-point functions
\cite{arkady}.} Hence in $d=2$, $A$ appears as the coefficient
appearing in the two-point function while in $d=4$, it appears in the
three-point function. However, for higher (even) dimensions, one must
examine higher $n$-point functions to determine $A$. It would be
natural/interesting to explore whether a similar story applies for odd
dimensions. In particular, it may seem natural to guess that $\ads$
might be identified using the two- and three-point functions for $d=3$.
Of course, the analysis above indicates that the coefficient of the
two-point function $\ct$ and $\ads$ are independent constants. Further
our definition \reef{adfun3} for $\ads$ in $d=3$ includes a
contribution coming from the four-dimensional Euler density in the
gravitational action. However, the latter is purely a topological term
in this case and so will not effect any equations of motion for
gravitational fluctuations. Hence the corresponding coupling constant
will not appear in any of the $n$-point functions in the boundary
theory suggesting that we cannot reproduce $\ads$ from these
correlators. Similar statements apply for our result \reef{adfun5} for
$\ads$ in $d=5$.

Above, we have shown that in our holographic models, $\ads$ cannot be
identified with $\ct$ or $C_S$. Of course, it could be true that, in
our models or in higher dimensional field theories in general, more
than one quantity obeys a c-theorem, \ie an inequality of the form in
eq.(\ref{magic2}). Following the discussion in section 2, one can
easily construct flow functions which yield $\ct$ or $C_S$ in the
AdS$_{d+1}$ vacua. However, examining their behaviour in general RG
flow geometries, one cannot readily establish that their radial
evolution is monotonic.

However, let us examine the behaviour of $C_S$ in a little more
detail.\footnote{We focus on $C_S$ since, as commented above, we don't
expect $\ct$ to obey a c-theorem in general.} To make progress, let us
consider Gauss-Bonnet (GB) gravity, \ie we set $\mu=0$ in section
\ref{one}. In the RG flows, the only quantity that changes is $\alpha$
and so it is more useful to express $C_S$ as
 \be
C_S= \frac{d+1}{d-1}\frac{\Gamma[d+1]}{\pi^{d/2}\Gamma[d/2]}\,
\frac{L^{d-1}}{\lp^{d-1}}X^{d-1}\quad{\rm with}\
X\equiv\frac{1}{\fin^{1/2}(1-\lambda\fin)}
 \labell{csfinal9}
 \ee
where only the last factor $X^{d-1}$ evolves in the RG flows. Having
restricted our attention to GB gravity, eq.~\reef{cubic} simplifies to
a quadratic equation and we can solve for $\fin$
 \be
\fin=\frac{1-\sqrt{1-4\lambda\alpha}}{2\,\lambda}\,.
 \labell{fine}
 \ee
\FIGURE[t]{
\includegraphics[width=0.8\textwidth]{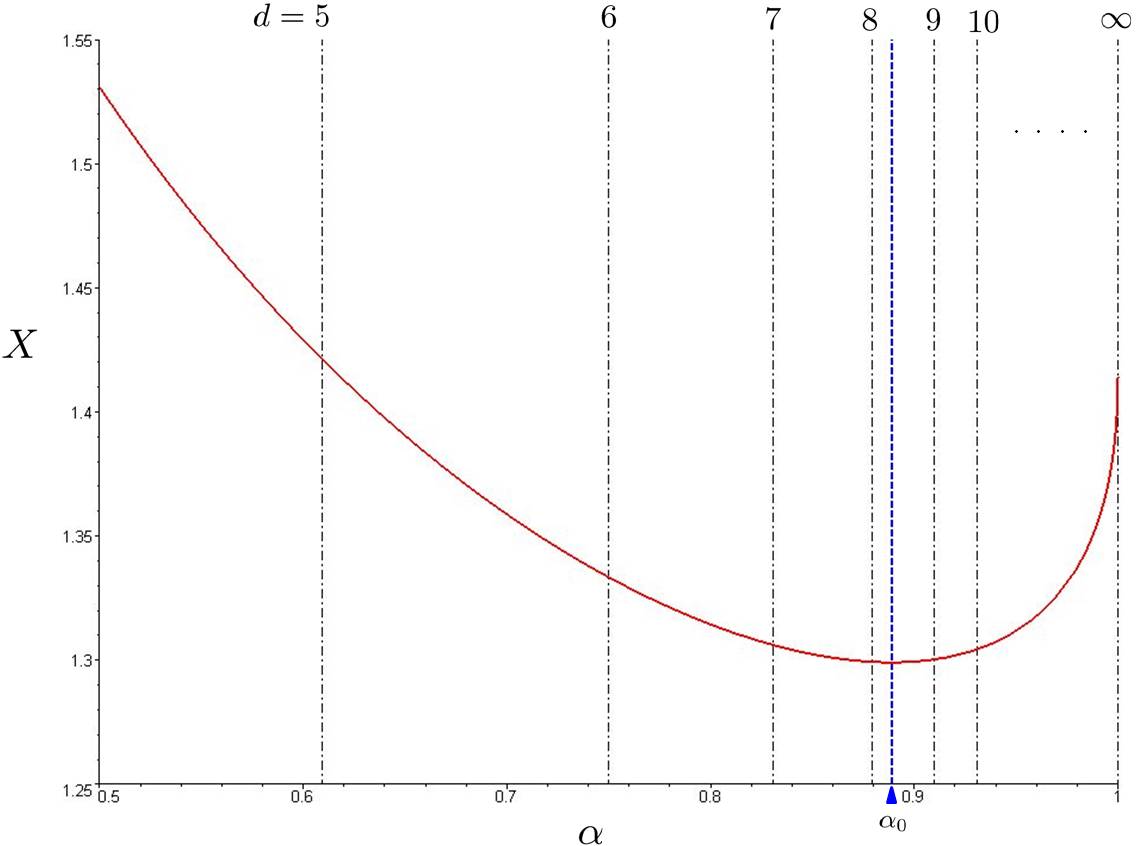}
\caption{The factor $X$ as a function of $\alpha$ (with $\lambda=1/4$).
The thick blue vertical dashed line indicates $\alpha_0$ in
eq.~\reef{alpha0}. From left to right, the thin black vertical dashed
lines indicate the critical value \reef{criticala} for $d=5,6,7,8,9,10$
and $\infty$. The corresponding line for $d=4$ would appear to beyond
the left boundary of the plot, \ie at $\alpha=0.36$. We see that for
$d\ge9$ there is a regime where flows could yield
$(C_S)_{UV}<(C_S)_{IR}$} \label{picture9}}

Using this result \reef{fine}, we plot the result for $X$ in
figure~\ref{picture9}, where we see that its slope changes from
negative to positive for `large' values of $\alpha$. In particular, it
is a straightforward show that the slope vanishes at
 \be
\alpha_0=\frac{2}{9\lambda}\,.
 \labell{alpha0}
 \ee
The significance of this observation is that in RG flows the value of
$\alpha$ always increases in going from a UV fixed point to an IR fixed
point. Hence any flows in which $\alpha\le\alpha_0$ will always satisfy
$(C_S)_{UV}>(C_S)_{IR}$ in this class of models. On the other hand,
flows which extend past the critical value $\alpha_0$ can violate this
inequality.

Now we recall that the analysis of \cite{consist,holoGB} shows that in
($d$+1)-dimensional GB theory with fixed $\alpha=1$ there is an upper
bound on the gravitational coupling $\lambda$, \ie
 \be
 \lambda\le\lambda_c=\frac{(d-2)(d-3)(d^2-d+6)}{4(d^2-3d+6)^2}\,.
 \labell{criticall}
 \ee
This ensures that the dual field theory was causal and has positive
energy fluxes. In the present framework, where $\lambda$ is fixed but
$\alpha$ varies, this translates into an upper bound on the latter
 \be
 \alpha\le\alpha_c=\frac{\lambda_c}{\lambda}\,.
 \labell{criticala}
 \ee
Hence we will not violate the desired c-theorem in this `physical'
regime as long as this critical value is smaller that $\alpha_0$. It
turns out that this inequality is satisfied for $d\le8$ but violated
for $d\ge9$, as illustrated in figure \ref{picture9}. Hence it seems
that we are able to violate the desired c-theorem in this class of
holographic models when the spacetime dimension is large.

Note that these violations occur very near the boundary
$\alpha=\alpha_\mt{max}=1/(4\lambda)$, beyond which the GB theory has
no AdS vacua (or any other maximally symmetric vacuum solutions). That
is, the roots of eq.~\reef{cubic} with $\mu=0$ are both complex, as can
be seen in eq.~\reef{fine}. We might note that as
$\alpha\to\alpha_\mt{max}$, the central charge $\ct$ also approaches
zero. Further, one finds that the ratio of shear viscosity to entropy
density can approach zero at this boundary \cite{fate}. These unusual
physical properties, as well as the fact that RG flows may yield
$(C_S)_{UV}<(C_S)_{IR}$, may indicate that the theory is actually
pathological in this regime, although we have no direct evidence of any
pathology at this point.

\section{Discussion} \label{discuss}

In sections \ref{one} and \ref{two}, we have examined RG flows for a
broad class of holographic models and found that these theories
naturally gave rise to a holographic c-theorem. Further, in sections
\ref{EE1} and \ref{EE2}, we showed that the quantity satisfying the
c-theorem, \ie satisfying eq.~\reef{beta3}, can be identified with the
coefficient of the universal contribution to the entanglement entropy
in a particular construction. Our results led us to make the following
general conjecture \cite{new}:
\begin{itemize}
\item {\it Placing a $d$-dimensional CFT on $S^{d-1}\times R$ and
    calculating the entanglement entropy of the ground state
    between two halves of the sphere, one finds a universal
    contribution: $S_{univ}\propto a^*_d$ (as detailed in
    eq.~\reef{unis2}). Then in RG flows between fixed points,
    $(a^*_d)_{\rm UV}\ge(a^*_d)_{\rm IR}$.}
\end{itemize}
In particular, our results lend credence to the idea that the universal
coefficients appearing in entanglement entropy play the role of central
charges in odd-dimensional CFT's. Hence our conjecture above provides a
framework to consider the c-theorem for quantum field theories in
spacetimes with either an odd or even number of dimensions.

For the case of even-dimensional boundary theories, we were able to
show that the universal coefficient that we identified using the
entanglement entropy was precisely the coefficient of the A-type trace
anomaly \reef{trace}. In fact, in section \ref{EEcft}, we were able to
show that for any even-dimensional CFT, the universal coefficient
appearing the entanglement entropy is precisely the A-type central
charge, without any reference to holography. Therefore our conjecture
above coincides precisely with Cardy's proposal \cite{cardy} for any
even $d$ and any evidence for Cardy's conjecture also supports the
present conjecture. However, the present results also frame Cardy's
conjecture in the context of entanglement entropy. This new perspective
may provide some useful insights towards constructing a general proof
of the c-theorem in higher dimensions. It is quite remarkable, that
after more than twenty years, no counter-example is known but a general
proof of Cardy's proposal is still lacking.

Of course, entanglement entropy has previously been considered in the
context of RG flows and c-theorems. In particular, \cite{casini}
provided a complementary proof of Zamalodchikov's c-theorem \cite{zam}
in $d=2$ based on considerations of entanglement entropy, using Lorentz
symmetry and the strong subadditivity. One obstacle to generalizing
their analysis to higher dimensions was that it was unclear what
geometry to specify in calculating the entanglement entropy
\cite{casini}. In our conjecture, we specify a very particular
geometry. This prescription for the geometry was essential to finding
$S_{univ}\propto A$ in even $d$, whereas for a general surface, this
coefficient in the entanglement entropy will be given by some linear
combination of all of the central charges appearing in the trace
anomaly \cite{taka2,solo} -- as described in section \ref{EEcft}. Hence
our construction may provide a starting point to extend the results of
\cite{casini} to higher $d$. It would, of course, be interesting to
identify other geometries for which the universal entanglement entropy
is proportional to $\ads$. At least for even $d$, one can show that the
same universal coefficient appears when the entanglement entropy is
calculated across a spherical boundary in flat space \cite{casini9}.
That is, the central charge $A$ controls the entanglement entropy in
this geometry as well -- a result that was already known for general
four-dimensional CFT's \cite{solo} and a massless conformally coupled
scalar for any even $d$ \cite{free,free2}.\footnote{For comparison
purposes, the appropriate coefficient for the trace anomaly of a
conformally coupled scalar may be found in \cite{cappelli}. We further
note that \cite{free2} also provides a result for a four-dimensional
Maxwell field which disagrees with the general result \cite{solo} that
$S_{univ}\propto A$. However, we expect that this can be corrected by
accounting for certain subtleties in the vector calculation, as
elaborated in \cite{subtle}.}

Further evidence for Cardy's proposed c-theorem and our conjecture
above can be found by examining RG flows induced by double-trace
operators \cite{aller,aller2}. Of course, a greater challenge is to
find evidence for the conjectured c-theorem outside of a holographic
framework. As explained above, for even $d$, this conjecture coincides
with Cardy's proposal \cite{cardy} and so implicitly any evidence for
the latter also supports the present conjecture. Hence it is more
interesting to consider the case of odd $d$. We can infer that the
entanglement entropy decreases along RG flows for certain known
examples in $d=3$ (and 2) \cite{vidal2,subir,fradkin}. However, we
cannot say that these examples provide direct support of our conjecture
primarily because the entanglement entropy is not calculated for the
geometry specified here.

Here we must acknowledge a technical point with regards to odd
$d$.\footnote{We are grateful to Horacio Casini, Adam Schwimmer, Misha
Smolkin and Stefan Theisen for conversations on this point.} Following
\cite{taka2}, we have identified the universal term in the entanglement
entropy as the constant term appearing in the expansion of
eq.~\reef{stotal1} for large $\rho_{max}$ or for a small cut-off
$\delta=\tL/\rho_{max}$. The universality of this constant contribution
to the entanglement entropy is well-established for a wide variety of
$d=3$ conformal quantum critical systems \cite{fradkin}, as well as
certain three-dimensional (gapped) topological phases \cite{wenx}.
However, in our calculations, it seems that we could replace the
cut-off with $\delta' =\delta +\delta^2/\ell$ where $\ell$ is some
macroscopic scale. Such a redefinition of the cut-off would modify the
above expansion and in particular change the constant contribution. One
objection to such a redefinition of the cut-off should be that at a
conformal fixed point, there is no natural macroscopic scale that could
play the role of $\ell$. In other words, the only macroscopic length
scales would seem to be defined by the geometry of the background or
the region in the entanglement entropy calculation, but it seems
unnatural that the cut-off would be dependent on these scales. A
further observation is that if one could restrict the allowed
redefinitions to be of the form $\delta' =\delta\times f(\delta/\ell)$
where $f$ is even, then the form of the expansion for the entanglement
entropy with odd $d$ would in fact leave the constant contribution
invariant. As a final comment, let us add that in field theory, we are
used to the idea that the precise value of, say, a gauge coupling
constant is scheme dependent. However, once a scheme is fixed, there is
no problem in defining the behaviour of the coupling under
renormalization group flow. It could be that a similar resolution
applies here for our c-theorem for odd dimensions.

Unfortunately we have no conclusive arguments in regard to the above
issues at present. Hence we may have to refine further our
characterization of $\ads$ in terms of entanglement entropy described
in sections \ref{EE1} and \ref{EE2}. For example, we expect that these
issues can be circumvented by considering the mutual information with
an appropriate construction \cite{casini9}. The mutual information is a
combination of entanglement entropies given two spatial subsystems:
$I(A,B) = S(A) + S(B)- S(A\cup B)$ -- for example, see
\cite{casini55,swingle}. In particular, with separate regions, the
subtraction ensures that $I(A,B)$ is free of divergences and any
ambiguities.

We provided two holographic calculations of the entanglement entropy
for the specific geometry described above. First in section \ref{EE1},
the entanglement entropy is calculated by relating it to the thermal
entropy of the CFT on the hyperbolic plane and at a particular
temperature. In section \ref{EE2}, we constructed a holographic
calculation based on the replica trick and we reproduced the same
results as in the previous section. Both of these calculations are
distinct from the standard holographic calculations of entanglement
entropy \cite{taka1,taka2}, which involves finding the area of a
minimal surface in the bulk with ends on the appropriate surface in the
boundary. In particular, the latter proposal only applies when gravity
in the bulk is described by Einstein's theory and so it cannot be used
with gravitational theories with higher curvature interactions. Here we
emphasize that the analysis in sections \ref{EE1} and \ref{EE2} holds
for a completely general covariant gravity action and is not restricted
to the theories introduced in sections \ref{one} and \ref{two}. On the
other hand, if we set to zero the coupling constants controlling the
higher curvature interactions, the bulk theory reduces to Einstein
gravity and our results for the entanglement entropy would match those
found using the standard approach. We also note that at present, there
is no derivation for this standard proposal for holographic
entanglement entropy \cite{taka1,taka2}.\footnote{As emphasized in
\cite{headrick}, the derivation presented in \cite{furry} is flawed
and, for example, leads to incorrect results in holographic
calculations of Renyi entropies.} Hence our discussion in sections
\ref{EE1} and \ref{EE2} then provides the first such derivation, albeit
for a calculation of entanglement entropy in a particular geometry. In
this regard, our results put the standard proposal \cite{taka1,taka2}
on a firmer footing since we find agreement in the limit of Einstein
gravity. Our derivation in section \ref{EE1} can be further extended to
calculations of entanglement entropy across a general spherical
boundary \cite{casini9}. There has also been some other recent progress
in understanding holographic entanglement entropy with higher curvature
theories in the bulk \cite{friends2}.

An intuitive understanding of the c-theorem leads to an interpretation
that the c-function provides a measure of the number of degrees of
freedom of the underlying QFT. Hence the coefficient of the type-A
trace anomaly should play this role for CFT's with even $d$
\cite{cappelli}. Our discussion in section \ref{density} makes this
connection precise for holographic CFT's in eq.~\reef{Waldformula6}. Of
course, this result applies equally well for odd or even $d$ and so
$\ads$ still plays exactly the same role in counting the degrees of
freedom of the dual CFT in either case. We emphasize that this result
applies for any gravitational theory with a covariant action
\reef{covariantM}. It would be interesting to better understand the
connection of this role of counting degrees of freedom and the
appearance of $\ads$ as a universal coefficient in entanglement
entropy. In this regard, it is noteworthy that in the bulk analysis
both results rely on Wald's entropy formula \reef{Waldformula} for
black hole horizons. In any event, it is clear that the coefficient
$\ads$ will play an important role in general holographic theories.

At this point, we re-iterate that our results for holographic
entanglement entropy in section \ref{EE2} hold for any bulk theory with
a general covariant gravity action. In fact, it is not difficult to
extend the approach of section \ref{EE1} to a completely general
gravity theory as well \cite{casini9}. On the other hand, our results
with regards to holographic c-theorems required tuning the couplings in
our gravitational action \reef{action} in a particular way. Our
motivation here was that the resulting holographic model should be
physically reasonable, \ie the dual CFT should be unitary. The reader
may wonder why we limited our models not to include only higher
curvature interactions beyond the order curvature-cubed? In fact, there
is nothing special about this truncation which was made to simply
provide a relatively simple framework in which to calculate. It would
of course be interesting to confirm that our general approach extends
to higher curvature interactions beyond curvature-cubed terms. Some
progress in this direction has been made in \cite{miguel9x}. Again we
wish to emphasize that our gravity theories should be regarded as toy
models which allow us to explore of the role of higher curvature terms
in holography. These gravity actions do not obviously emerge from any
string theory calculations. One could consider a `perturbative' limit
of our analysis. That is, our results would still carry over in a limit
where $\lambda_i=O(\lp^2/L^2)$ and $\mu_i=O(\lp^4/L^4)$, as is done in
most string theory calculations. The only caveat then would be that one
expects that string theory to also generate higher derivative
interactions in the matter sector. Although it may be possible to
eliminate all or some of these with field redefinitions \cite{alex6},
in general it is not clear what would replace the null energy condition
in this context -- we return to this discussion below. A long-term goal
remains to better develop our ability to calculate within a string
theory framework in order study interesting holographic backgrounds
with large curvatures and the corresponding RG flows.

Our construction of physically reasonable holographic models in section
\ref{two} focussed on the idea that the boundary theory should be
unitary. That is, we tuned the coupling constants for the higher
curvature interactions to ensure the the linearized graviton equations
are only second order in derivatives. This tuning eliminates the
appearance of ghost modes in the bulk theory and of non-unitary
operators in the boundary QFT. That such tuning is possible for AdS
backgrounds was originally observed for quasi-topological gravity in
\cite{old1} and later in lower dimensional theories as well
\cite{sinha,paulos,aninda9} -- see also the discussion in \cite{ssc}.
Again, this approach can also be extended to higher dimensions with
higher curvature interactions beyond curvature-cubed terms
\cite{miguel9x}. For the present analysis, we demanded that this
property apply for general RG flows, \ie for fluctuations around the
general metric \reef{metric}, and restricted our general theory
\reef{action} with eleven coupling constants down to a model with only
four independent couplings. Having imposed the unitarity constraints
(\ref{2dera}--\ref{2dere}), the resulting model automatically satisfied
a holographic c-theorem. It would be interesting to see if this result
extends to more general theories, \eg with higher curvature terms
beyond $R^3$.\footnote{Recall that no three-dimensional gravity theory
can constructed with only curvature-squared and curvature-cubed
interactions which simultaneously satisfies a c-theorem and is
`unitary' \cite{sinha}.}

In this analysis, a key property of the RG flow geometries
\reef{metric} seems to be that they are conformally flat. Of course,
there are many physically interesting geometries (\eg general AdS black
hole solutions) which will not share this properties and in these
backgrounds, higher derivative terms will reappear in the linearized
graviton equations.  For example, in quasi-topological gravity, the
four-derivative contributions can be expressed as couplings with the
background Weyl curvature \cite{old1}
 \beq
W^{ cd ef}\,h_{ de; cf\left( a b\right)} +2\left(\Box h^{
c}{}_{\left(\right. a }\right){}^{\!; de}W_{| cd e| b\left.\right)}
+2\,\Box^2h^{ cd }W_{ c a  d b}+g_{ a  b}\left(\Box h_{ cd }\right)_{;e
f}W^{ ce d f}\,.\labell{fourder}
  \eeq
For simplicity, we are only considering transverse traceless modes here
(\ie $\nabla^a h_{a b} =0=h^a{}_a$).\footnote{In eq.~\reef{fourder}, we
have also adopted the standard notation: $T_{(ab)}=\frac{1}{2}\left(
T_{ab} + T_{ba}\right)$.} Hence in a background where $W_{abcd}\ne0$,
the analysis of the linearized fluctuations becomes far more
complicated, \eg see \cite{old2}.

Let us make two observations at this point: First of all, any
asymptotically AdS background describing a scenario where the boundary
theory is Poincar\'e invariant will have a form which matches the
metric ansatz \reef{metric}, which we used to consider general RG
flows. Hence our analysis in section \ref{two} ensures that no higher
derivatives arise for such a Poincar\'e invariant background. The
implication is then that higher derivative terms can only arise when
the boundary theory or boundary state is not Lorentz invariant.
Therefore the appearance of such terms need not immediately imply the
existence of ghost modes. Of course, theories with a Lifshitz symmetry
provide simple examples where higher derivative equations of motion do
not imply the appearance of ghosts (\eg $(\partial_t^2+\kappa
\nabla^4)\phi=0$).

Another interesting aspect of the higher derivative terms, such as
those illustrated in eq.~\reef{fourder}, is that they will vanish in
the asymptotic region with AdS boundary conditions, since the Weyl
curvature will vanish there. Hence any new modes associated with these
terms (which are potentially unstable or ghost-like) will be `confined'
to the interior of the geometry. Hence it seems that such modes will
not be associated with the appearance of new operators in the boundary
theory. Rather such modes would reflect some complicated new infrared
dynamics which would be insensitive to the ultraviolet details of the
theory. We recall that despite the fact that the full equations of
motion of quasi-topological gravity are fourth order in derivatives,
the most general static black hole solutions are characterized by a
single integration constant \cite{old1} -- just as in Einstein (or
Lovelock) gravity. This result is likely related to this idea that the
four-derivative contributions in the linearized equations have no
effect near the asymptotic boundary. To properly understand these
issues, it seems that one should analyse the asymptotic geometry of a
general solution of our `unitary' higher curvature theories using a
Fefferman-Graham-like expansion \cite{feffer}, \eg following
\cite{sken}.

Our final comment in this regard is that our restriction to
second-order linearized equations is sufficient but not necessary to
ensure unitarity of the boundary theory. In particular, consider
restricting the action \reef{action} to contain only the terms
containing the Ricci scalar. The resulting equations of motion are
indeed fourth order in derivatives but it is well know that such an
$f(R)$ theory is equivalent to a theory of Einstein gravity coupled to
a `ordinary' scalar field, \eg see \cite{fofr}. If one considers the
fourth order linearized equations, one would find that the only extra
modes correspond to a massive scalar whose propagator comes with a
positive sign, \ie these modes are not ghost-like -- for example, see
\cite{sugra,japan}. Hence in the AdS/CFT framework, the metric
fluctuations couple to a new unitary scalar operator, as well as the
stress tensor, in the boundary theory. It would be interesting to do a
more general analysis of the gravity equations of motion for the action
\reef{action} to understand precisely which of the constraints in
section \ref{two} can be relaxed while still preserving unitarity in
boundary theory.

In constructing our holographic models in sections \ref{one} and
\ref{two}, we have introduced an unconventional gravity theory, \ie a
higher curvature theory, and a conventional matter theory. Both sectors
were constrained in different ways. That is, as we have just discussed,
the gravitation couplings were tuned to ensure that no non-unitary
operators appear in the dual QFT, while the matter sector was required
to satisfy the null energy condition. One interesting question is to
better understand the holographic interpretation of the null energy
condition in terms of the boundary theory. However, it would also be
more interesting to consider more general couplings between the
gravitational and matter sectors. For example, if the matter sector was
simply a scalar with an interesting potential, one can imagine that
rather than having constant gravitational couplings that these would be
replaced by functions of the scalar field. That is, we would introduce
interactions of the form $\W_1(\phi)\,R^2$. Enforcing `unitarity'
constraints extends to the generalized theory in an obvious way, but
will now involve linearized equations of motion for both the metric and
scalar fields. However, it is not clear what constraints should replace
the null energy condition to ensure a holographic c-theorem. Some
preliminary analysis of RG flows in the presence of such generalized
couplings are presented in appendix \ref{ageneral}. A full
understanding of these issues seems to be a challenging problem which
we leave to future work.

Another interesting direction for the future would be to investigate
holographic c-theorems for models with non-relativistic symmetries.
Models with Schr\"odinger \cite{schro} and Lifshitz \cite{liff}
symmetries have been studied and further the contribution of higher
curvature terms has also been considered in this context \cite{next}.
Here we propose to consider RG flows between different vacua of such a
model. In particular, it would be interesting to consider flows where
the symmetry group changes between the UV and IR fixed points. Again an
initial investigation of some of these questions in given in appendix
\ref{anonrel} but a full examination is left to the future.

\vskip 2cm

\noindent {\bf Acknowledgments:} We thank John Cardy, Horacio Casini,
Sumit Das, Stuart Dowker, Ben Freivogel, Rajesh Gopakumar, Matt
Headrick, Janet Hung, Dileep Jatkar, Andr\'e LeClair, Shiraz Minwalla,
Miguel Paulos, Adam Schwimmer, Al Shapere, Misha Smolkin, Brian
Swingle, Lenny Susskind, Yuji Tachikawa, Stefan Theisen and Toby
Wiseman for useful discussions. AS thanks the Physics Departments of
the University of Kentucky and Princeton University where part of this
work was presented. Research at Perimeter Institute is supported by the
Government of Canada through Industry Canada and by the Province of
Ontario through the Ministry of Research \& Innovation. RCM also
acknowledges support from an NSERC Discovery grant and funding from the
Canadian Institute for Advanced Research. RCM thanks the Galileo
Galilei Institute for Theoretical Physics for hospitality and the INFN
for partial support during the completion of this work.

\appendix

\section{RG flows} \label{aflow}

In this appendix, we illustrate the holographic RG flows in more
detail. For concreteness, we consider the quasi-topological theory with
$d=4$ as discussed in section \ref{one}. Recall that the curvature in
the AdS$_5$ vacua is set by
 \be
\frac{1}{\tilde L^2}=\frac{\fin}{L^2}\,,\quad
\alpha=\fin-\lambda\fin^2-\mu\fin^3\,.
 \labell{alphaeq}
 \ee
As explained in section \ref{one}, we can imagine that the gravity
theory is coupled to, \eg a scalar field with an interesting potential
which yields various stationary points. These different stationary
points will be distinguished by different values of the (negative)
cosmological constant, \ie they yield different values for the
parameter $\alpha$ above. As discussed, we consider the root of
eq.~(\ref{alphaeq}) that is smoothly connected to $\fin=\alpha$ in the
limit $\lambda,\mu\rightarrow 0$.

At each of the stationary points, there are three dimensionless
parameters which characterize the gravitational theory, the couplings
$\lambda$ and $\mu$ as well as the ratio $\tL/\lp$. Similarly, the dual
CFT at the corresponding fixed points can be characterized in terms of
three parameters defining the three-point function of the stress tensor
\cite{osborn}. Alternatively, these parameters define various other
constants which define the dual CFT, \eg the central charge $a$. Two
other convenient constants are $t_2$ and $t_4$, which arise in certain
gedanken experiments described in \cite{hofmal}. In our holographic
framework, these two parameters are given by
\cite{old2}:\footnote{These formulae were originally derived in
\cite{old2} with $\alpha=1$ but the results remain unchanged when one
allows $\alpha$ to take on different values in eq.~\reef{alphaeq}.}
 \be
t_2=\frac{24 \fin (\lambda-87 \fin \mu)}{1-2 \lambda \fin-3\mu
\fin^2}\,, \qquad t_4=\frac{3780 \mu \fin^2}{1-2 \lambda \fin-3\mu
\fin^2}\,.
 \labell{tt24}
 \ee
In figure \ref{picture1}, we plot $t_4$ against $t_2$ for various
models. The contours are generated by varying $\alpha$ while keeping
the couplings $\lambda$ and $\mu$ fixed. These lines should not be
thought of as literal RG flows in parameter space, \ie $t_2$ and $t_4$
are only defined at the fixed points. Rather any RG flow in this model
will connect two fixed points on the same contour. Note that the
origin, \ie $t_2=0=t_4$, corresponds to Einstein gravity and the RG
flows generally move out into the plane away from this point. Note
however that the RG flows never cross the lines $\mu=0$ or $\lambda=0$,
which correspond to $t_4=0$ and $t_2+\frac{58}{105}t_4=0$,
respectively. The gedanken experiments defining $t_2$ and $t_4$ also
establish that the corresponding CFT will only be consistent within a
(small) region around the origin \cite{hofmal}. The boundaries of this
physical region are indicated as the `flux constraints' in red in
figure \ref{picture1}. It would be straightforward to construct the
holographic RG flows which pass out through these boundaries. Hence it
appears the RG flows are not a refined enough probe of the holographic
model to detect the inconsistencies arising in the unphysical region.
However, we note that there can be no such flows which enter into the
physical region from outside.
\FIGURE[t]{
\includegraphics[width=0.9\textwidth]{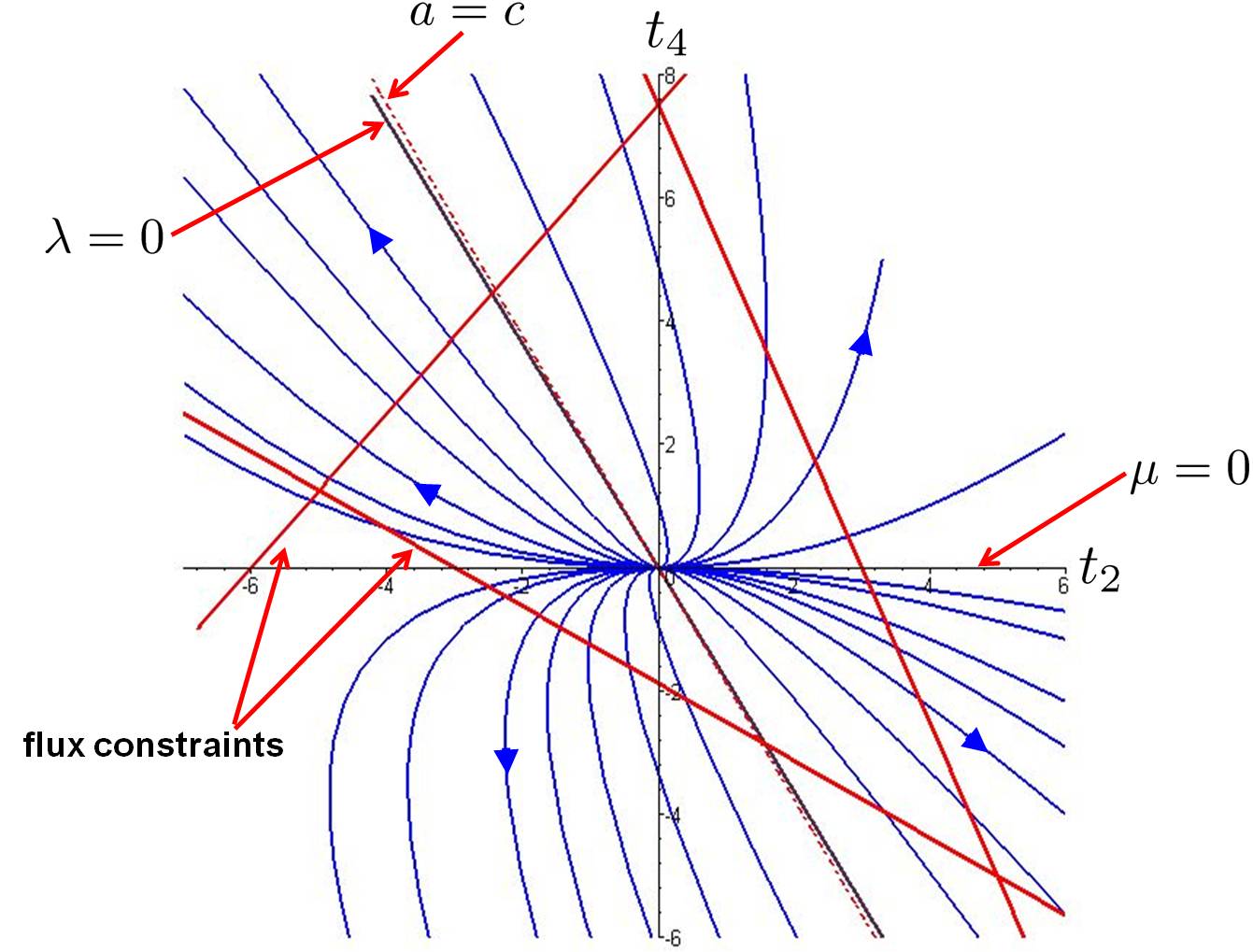}
\caption{RG flows in the $(t_2,t_4)$ plane. Holographic RG flows (in
the model described in the main text) connect fixed points on the same
contour. The arrows indicate the direction of the flows towards the
infrared. The black line indicates $\lambda=0$. No trajectories cross
this line or the line $\mu=0$. The dashed red line indicates $a=c$ (or
$t_2+\frac{8}{15}t_4=0$) and in principle, it is possible to construct
holographic RG flows which cross this line.} \label{picture1}}

To conclude this discussion, however, we must say that we cannot be
certain at this point which, if any, of the detailed aspects of the RG
flows noted above can be regarded as universal properties. That is,
which properties apply outside of the particular class of holographic
models based on quasi-topological gravity \reef{ActD}.

\section{More general couplings} \label{ageneral}

In this section, we initiate the study of c-theorems with more general
couplings between the gravitational sector and the matter fields in the
bulk. To be concrete, let us consider the following action
 \be
I=\frac {1}{2\lp^{3}}\int d^{5}x \sqrt{-g} \left[\frac{12}{L^2}{\cal
V}(\phi)+R-\frac{1}{2}\partial_\mu\phi
\partial^\mu \phi+ \frac{L^2}{2} \W(\phi)\chi_4\right]
\labell{action1}
 \ee
where $\chi_4$ is the combination of curvature-squared terms
corresponding to the four-dimensional Euler density given in
eq.~\reef{GBterm}. Above, we have explicitly coupled gravity to a
scalar field with a standard kinetic term and a potential
$V(\phi)=-12\,{\cal V}(\phi)/L^2$. The new feature of this theory is
that we have also introduced a scalar coupling $\W(\phi)$ to the
curvature-squared interaction. We might note that despite this new
coupling, the equations of motion derived from the above action
\reef{action1} are purely second order in derivatives for any arbitrary
background \cite{newe} and hence the dual theory is `unitary',
according to the discussion given in section \ref{two}.

As in the main text, we assume that this action yields several
stationary points where the scalar is constant, \ie $\phi=\phi_i$, and
the bulk geometry is simply AdS$_5$. Recall that the latter corresponds
to $A(r)=r/\tL$ in flow metric \reef{metric}. If we again adopt the
notation that the AdS curvature scale is written as $\tL^2=L^2/\fin$,
then eq.~\reef{cubic} is replaced by
 \be
\V(\phi_i)=\fin-\W(\phi_i)\,\fin^2\,.
 \labell{new8}
 \ee
Note that since $\chi_4$ is nonvanishing in the AdS$_5$ background, the
stationary points for the scalar field depends on both $\V$ and $\W$,
as well as $\fin$ (\ie the curvature). That is, the critical values
$\phi_i$ must satisfy
 \be
\left[\frac{\delta\V}{\delta\phi}+10\fin^2\,
\frac{\delta\W}{\delta\phi}\right]_{\phi=\phi_i}=0\,.
 \labell{new7}
 \ee
These equations determining the vacuum solutions already illustrate
that there is no clean separation between the matter and the gravity
sectors with this new action \reef{action1}.

Further, given the lack of such a separation, there is no clear role of
the standard null energy condition nor is it clear what rule might
replace this condition. Hence we must look elsewhere to construct a
flow function for the new theory. Motivated by a possible connection
between the flow function and Wald's formula \reef{Waldformula}, we
consider
 \be
a(r)=\frac{\pi^2}{\lp^3 A'(r)^3}\left[ 1-6 L^2 A'(r)^2 \W(\phi)
\right]\,.\labell{new6}
 \ee
That is, we have constructed this function to satisfy
eq.~\reef{waldtype}. Now, using the equations of motion, we find
 \be
a'(r)=\frac{\pi^2}{\lp^3 A'(r)^4}\big[\phi'(r)^2-6 L^2 \partial_r
(A'(r)^2\partial_r \W)\big]\,. \labell{new9}
 \ee
In main text, we would have had $\W$ being simply a constant coupling
in which case the second term vanishes making the right-hand side
clearly positive and hence we would recover the monotonic flow
discussed there. However, in eq.~\reef{new9}, the right-hand side does
not have a definite sign. A necessary and sufficient condition for
$a(r)$ to grow monotonically is
 \be
6 L^2\partial_r (A'(r)^2 \partial_r \W)\leq \phi'(r)^2\,.
 \labell{nece}
 \ee
One might also consider a simpler sufficient (but not necessary)
condition
 \be
\partial_r (A'(r)^2 \partial_r \W)\leq 0\,. \labell{suff}
 \ee
However, if we demand that the end-points of the RG flow are CFTs, \ie
the bulk geometry is AdS$_5$, then $\phi'$ vanishes there and so we
also have $\partial_r \W=\delta \W/\delta\phi\,\phi'=0$ at both end
points. Of course, this means that if derivative in eq.~\reef{suff} is
anywhere negative, this must be compensated by it being positive
elsewhere along the flow. Therefore the only circumstance to in which
eq.~\reef{suff} is satisfied is if $\W$ is constant. It may be that a
more detailed investigation of RG flows for this theory would reveal
that in fact, the condition (\ref{nece}) is always satisfied, at least
if certain conditions are imposed on $\W$ and $\V$. One could start
such an investigation by constructing explicit flow solutions with
concrete choices of $\V$ and $\W$.

If eq.~\reef{nece} is satisfied for a broad class of models, we expect
that there should a better choice of the flow function for which the
radial derivative becomes manifestly positive. We did try to modify our
expression \reef{new6} above in various ways but we were unable to find
a new flow function which clearly grows monotonically in the radial
direction. One such attempt was to consider the following function
 \be
{\tilde a}(r)=\frac{\pi^2}{\lp^3 A'^3}\left[ 1-6 L^2 A^{\prime\,2}\,
\W(\phi)-6L^2A'\,\frac{\delta \W}{\delta\phi}\,\phi'
\right]\,.\labell{new6a}
 \ee
Note that the new term added above will vanish at the AdS$_5$ fixed
points and so the new flow function and that in eq.~\reef{new6} will
yield identical results when evaluated at the fixed points. In this
case, using the equations of motion, the radial derivative becomes
 \be
{\tilde a}'(r)=\frac{\pi^2}{\lp^3 A'^4}\left[ \phi^{\prime\,2}+12 L^2
\frac{\delta \W}{\delta\phi}\,\phi'\,(A^{\prime\,2})'\right]\,.
\labell{new9a}
 \ee
Unfortunately the result on the right-hand side is again not obviously
positive.

It would also be interesting to examine the RG flows of these
generalized holographic models along the lines discussed in the
previous appendix.

\section{Non-relativistic $c$-theorems} \label{anonrel}

There has been some interest in holographic models for non-relativistic
CFT's, in particular, with Schr\"odinger symmetry \cite{schro}. It is
relatively easy to extend our construction of a holographic $c$-theorem
to such a framework but the following should only be regarded as an
initial step in this direction. Let us consider Gauss-Bonnet gravity in
five bulk dimensions, \ie eq.~\reef{ActD} with $d=4$ and $\mu=0$. With
the non-relativistic Schr\"odinger symmetry in mind, we consider the
following metric ansatz
 \be
ds^2=dr^2-e^{2 A(r)}dt^2+ 2 e^{2B(r)} dt
dx+e^{2B(r)}dy^2+e^{2B(r)}dz^2\,.
 \ee
The usual holographic Schr\"odinger geometry \cite{schro} is recovered
with $A(r)= z\, r /L$ and $B(r)\sim r/L$, where $z$ is the dynamical
exponent.

Now let us consider holographic RG flows in this context. If we assume
that $T_t^t-T_r^r\leq 0$ or equivalently $T_{tx}e^{-2 B}-T_{rr}\leq 0$
then a simple c-theorem can be easily proved. Note that this condition
does {\it not} arise from the null energy condition unless $T_{tt}=0$.
In this case, the equations of motion yield
 \be
3 B''(r) (1-2\lambda L^2 B'(r)^2)=T_t^t-T_r^r\,,
 \ee
and so that the flow function
 \be
a(r)=\frac{1}{B'(r)^3}(1-6\lambda L^2 B'(r)^2)\,,
 \ee
grows monotonically with radius. Alternatively, of course, we have that
$a(r)$ is monotonically decreasing along RG flows. Using the null
energy condition, we can make one further useful
statement.\footnote{The null-energy condition has been considered
before in \cite{hoyos} in the context of non-relativistic symmetries.
There it was shown that with a Lifshitz symmetry, $z\leq 1$ is
necessary for this condition to hold.} Consider the following null
vector in the ($t,x$)-plane: $v^a=(0,e^{-A},1/2 e^{-2B+A},0,0)$. The
null energy condition then requires
 \be
T_{tt}e^{-2A} +T_{tx}e^{-2B}+\frac{1}{4}T_{xx}e^{-4B+2A}\geq 0\,.
 \ee
Now at the end points of the RG flow, we expect $A(r)$ and $B(r)$ take
the form noted above for a Schr\"odinger geometry. Using this, we
derived the following inequality for the dynamical exponents at the UV
and IR fixed points:
 \be (z_\mt{UV}^2-1)(1-2\lambda \frac{L^2}{\tL_\mt{UV}^2})\geq
(z_\mt{IR}^2-1)(1-2\lambda
\frac{L^2}{\tL_\mt{IR}^2})\frac{\tL_\mt{UV}^4}{\tL_\mt{IR}^4}\,.
 \labell{junk}
 \ee
For $\lambda=0$, \ie Einstein gravity in the bulk, if we assume that
$\tL_\mt{UV}>\tL_\mt{IR}$, then this inequality demands that $z_\mt{UV}
\ge z_\mt{IR}$. That is, the dynamical exponent is always smaller at
the infrared fixed point. It would be interesting to have a better
insight into this inequality \reef{junk} when $\lambda\ne0$ and also to
investigate these RG flows in more detail.

\end{document}